\documentclass[10pt,journal,compsoc]{IEEEtran}
\ifCLASSOPTIONcompsoc
  \usepackage[nocompress]{cite}
\else
  \usepackage{cite}
\fi
\usepackage{xcolor}
\usepackage[pdftex]{graphicx}
\usepackage{amsmath}
\usepackage{multirow}
\usepackage{fourier} 
\usepackage{array}
\usepackage{makecell}
\usepackage{algorithm}
\usepackage{algorithmic}
\usepackage{hyperref}
\usepackage{booktabs}
\usepackage{multirow}
\usepackage{longtable}
\usepackage{enumitem}
\usepackage{pifont}
\usepackage{tabularx}
\usepackage{float}
\usepackage{subfig}

\newcommand{\fix}[1]{\textcolor{red}{\textbf{\textit{#1}}}}
\newcommand{\boit}[1]{\textbf{\textit{#1}}}

\begin{document}
%
\title{A Guide to Particle Advection Performance}

\author{
Abhishek Yenpure$^1$,
Sudhanshu Sane$^2$,
Roba Binyahib$^3$,
David Pugmire$^4$,
Christoph Garth$^5$,
Hank Childs$^1$ 
\thanks{$^1$University of Oregon}
\thanks{$^2$Scientific Computing Institute, University of Utah}
\thanks{$^3$Intel}
\thanks{$^4$Oak Ridge National Laboratory}
\thanks{$^5$University of Kaiserslautern}
}
%
%

\markboth{Journal of \LaTeX\ Class Files,~Vol.~14, No.~8, August~2015}%
{Shell \MakeLowercase{\textit{et al.}}: Bare Demo of IEEEtran.cls for Computer Society Journals}
%



\IEEEtitleabstractindextext{%
\begin{abstract}
The performance of particle advection-based flow visualization techniques
is complex, since computational work can vary based on
many factors, including number of particles, duration, and mesh type.
Further, while many approaches have been introduced to optimize performance,
the efficacy of a given approach can be similarly complex.
In this work, we seek to establish a guide for particle advection performance by conducting a comprehensive survey of the area.
%
We begin by identifying the building blocks for particle advection
and establishing a simple cost model incorporating these building blocks. 
We then survey
existing optimizations for particle advection,
 using two high-level categories:
algorithmic optimizations and hardware efficiency.
The sub-categories of algorithmic optimizations include solvers, cell locators, I/O efficiency, and precomputation,
while the sub-categories of  hardware efficiency all involve parallelism:
shared-memory, distributed-memory, and hybrid.
Finally, we conclude the survey by identifying current gaps in 
particle advection performance, and in particular on 
achieving a workflow for predicting performance under various optimizations.

\end{abstract}

}

\maketitle

\IEEEdisplaynontitleabstractindextext

%
\IEEEpeerreviewmaketitle

\IEEEraisesectionheading{\section{Introduction}\label{sec:intro}}
%
%
%
%
\IEEEPARstart{F}{low} visualization techniques are used to
understand flow patterns and movement of fluids
in many fields, including
oceanography, aerodynamics, and electromagnetics.
%
Many flow visualization techniques operate by
placing massless particles at seed locations,
displacing those particles according to a vector field to form trajectories,
and then using those trajectories to create a renderable output.
Each of the trajectories are calculated via a series of ``advection steps,''
where each step advances a particle a short distance
by solving an ordinary differential equation.

Particle advection workloads can be quite diverse across different
flow visualization algorithms and grid types.
These workloads consist of many factors, including the number of particles,
duration of advection, velocity field evaluation, and analysis needed
for each advection step.
One particularly important aspect with respect to performance
 is the number of advection steps,
which derive from both the number of particles and their durations.
Many flow visualization techniques have numerous particles that
go for short durations, while many others have few particles that go for long
durations.
Some cases, like when analyzing flow in the ocean~\cite{Ozgokmen},
require numerous particles for long durations
and thus billions of advection steps (or more).
With respect to velocity field evaluation, uniform grids require only a few
operations, while unstructured grids require many more (for cell location
and interpolation).
%
In all,
the diverse nature of particle advection workloads makes it difficult
to reason about both the execution time for a given workload and
the potential improvement from a given optimization.

%

%

The main goal of this survey is to provide a guide for understanding
particle advection performance, including possible optimizations.
It does this in four parts.
First, Section~\ref{sec:background}
provides background on the building blocks for particle advection.
Second, Section~\ref{sec:costmodel}
introduces a cost model for particle advection performance,
to assist with reasoning about overall execution time and inform which aspects dominate runtime.
Third,
Section~\ref{sec:optimizations}
surveys algorithmic optimizations.
Fourth,
Section~\ref{sec:usinghwefficiently}
surveys approaches for utilizing hardware more efficiently,
with nearly all of these works utilizing parallelism.
Contrasting the latter two sections,
Section~\ref{sec:optimizations}
is about reducing the amount of work to perform,
while Section~\ref{sec:usinghwefficiently}
is about executing a fixed amount of work more quickly.
In terms of how to read this survey,
Sections~\ref{sec:optimizations} and~\ref{sec:usinghwefficiently}
can be read in any order.
In particular, readers interested in a particular algorithmic optimization
or technique for hardware efficiency can skip to the corresponding subsection.
That said,
readers new to this topic should start with Sections~\ref{sec:background}
and~\ref{sec:costmodel} to gain a basic understanding of particle advection
performance.

In terms of placing this survey into context with previously published
literature, we feel this is the first effort to provide a guide
to particle advection performance.
The closest work to our own is the survey on distributed-memory parallel
particle advection by Zhang and Yuan~\cite{zhang2018survey}.
Our survey is differentiated in two main ways.
First, our survey considers a broader context overall, 
i.e., it considers algorithmic optimizations and additional types of parallelism.
Second, our discussion of distributed-memory techniques
does a new summarization of workloads
and parallel characteristics (specifically Table~\ref{tab:distscale}),
and also has been updated to include works appearing since their publication.
There also have been many other excellent surveys involving flow visualization and particle advection:
feature extraction and tracking~\cite{post2003state},
dense and texture-based techniques~\cite{laramee2004state},
topology-based flow techniques~\cite{laramee2007topology} and a subsequent survey 
focusing on topology for unsteady flow~\cite{pobitzer2011state},
integration-based, geometric flow visualization~\cite{mcloughlin2010over},
and
seed placement and streamline selection~\cite{sane2020survey}.
Our survey complements these existing surveys ---
while some of these works consider aspects
of performance within their individual focal point,
none of the surveys endeavor to provide a guide to particle advection performance.
%
%

\section{Particle Advection Background}
\label{sec:background}

\begin{figure*}[th]
\centering
\includegraphics[width=0.625\linewidth]{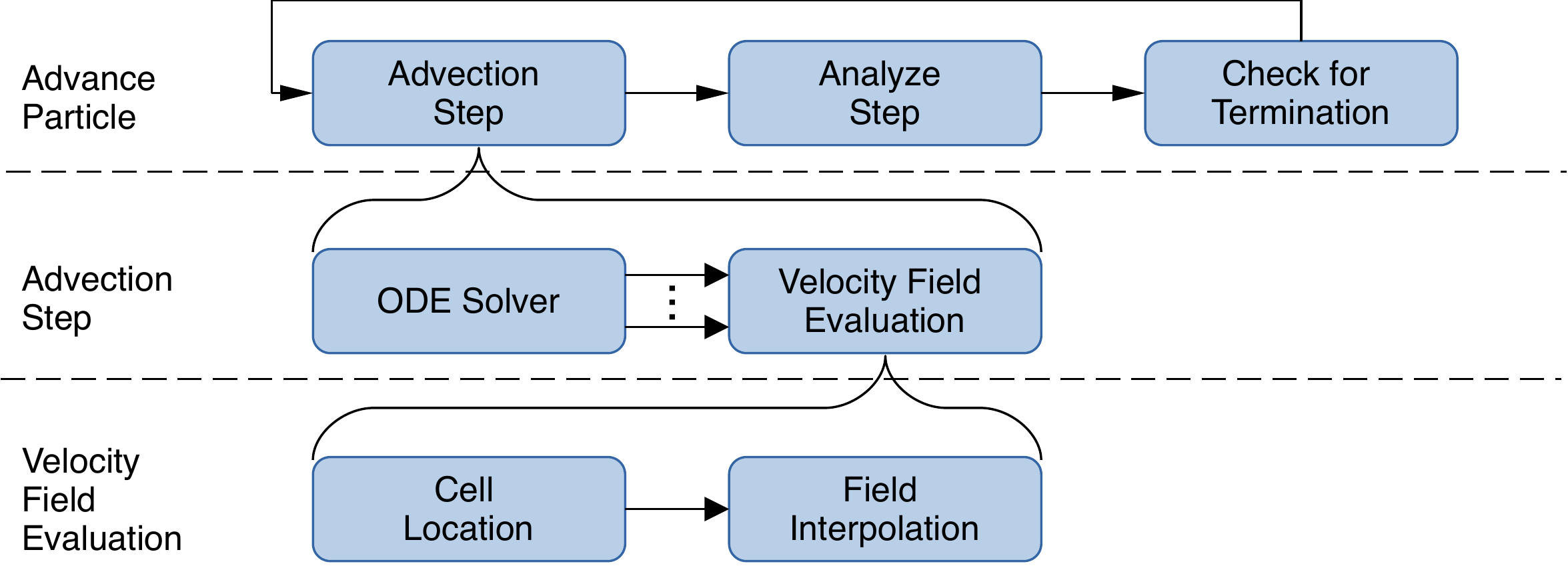}
\caption{\label{fig:pa_diagram}
Organization of the components for a particle advection-based flow visualization algorithm.
The components are arranged in three rows in decreasing levels of granularity from top to bottom.
In other words,
the components at the bottom are building blocks for the components at higher levels.
The top row shows components that define the movement and analysis of a particle.
The loop in the top row indicates its components are executed repeatedly
until the particle is terminated.
The middle row shows components that define a single step of advection.
The arrows with the ellipsis from ODE solver to velocity field evaluation
are meant to indicate that an ODE solver needs to evaluate the velocity field multiple times.
Each velocity field evaluation takes as input
a spatial location and possibly a time, and returns the velocity
at the corresponding location (and time).
The frequently-used Runge-Kutta 4 ODE solver requires
four such velocity field evaluations.
Finally, as depicted in the bottom row, each velocity field evaluation
requires first locating which cell in the mesh contains
the desired spatial location
and then interpolating the velocity field to the desired location.
}
\end{figure*}

Flow visualization algorithms perform three general operations:
\begin{itemize}
\item \boit{Seed Particles:} defines the initial placement of particles.
\item \boit{Advance Particles:} defines how the particles are displaced and analyzed.
\item \boit{Construct Output:}
constructs the final output of the flow visualization algorithm,
which may be a renderable form,
something quantitative in nature, etc.
\end{itemize}
These three operations often happen in sequence, but in some forms they
happen in an overlapping fashion (i.e., seed, advance, seed more, advance more, etc.)

Our organization, which is illustrated in Figure \ref{fig:pa_diagram},
 focuses on the ``advance particles'' portion of
flow visualization algorithms.
It divides the components into three levels of granularity.

The ``top'' level of our organization considers the process of advancing a single particle.
It is divided into three components:
\begin{itemize}
\item \boit{Advection Step:} advances a particle to its next position.
\item \boit{Analyze Step:} analyzes the advection step that was just taken.  
\item \boit{Check for Termination:} determines whether a particle should be terminated.
\end{itemize}
The process of advancing a particle involves three phases that
are applied repeatedly.
The first phase is
to displace a particle from its current location to a new location.
Such displacements are referred to as particle advection steps.
%
%
The second phase is to analyze a step.
The specifics of the analysis vary by flow visualization algorithm,
and could be as simple as storing the particle's new location in memory
or could involve more computation.
The third phase is to
check if the particle meets the termination criteria.
Similar to the second phase,
flow visualization algorithms define specific criteria for when to terminate a particle.
Finally, if the particle is not terminated, then these three phases are
repeated in a loop until the termination criteria
are reached.

The ``middle'' level of our organization
considers the process of completing a single step for a particle.
This level has two components:
\begin{itemize}
\item \boit{ODE Solver}:
calculates a particle's displacement to a new position by solving an ordinary diffential equation (ODE).
\item \boit{Velocity Field Evaluation:}
calculates the velocity value at a specific location by interpolating
within the located cell.
\end{itemize}
Thus, to calculate the velocity at a point P, cell location is first used to identify the cell C that contains P, and then velocity field interpolation is performed to calculate the velocity at P using information at the vertices of C. 

The ``bottom'' level of our organization considers
the process of velocity field evaluation.
This level also has two components:
\begin{itemize}
\item \boit{Cell Location:} locates the cell that contains some location.
\item \boit{Field Interpolation:}
calculates velocity field at a specific location via interpolation of
surrounding velocity values.
\end{itemize}
In terms of a relationship, to calculate velocity at some point $P$,
first cell location is used to identify the cell $C$ that contains $P$,
and then
velocity field interpolation  is used to calculate the velocity
at $P$ using $C$'s information.

\begin{figure*}[th]
\centering
\subfloat[streamlines]{\includegraphics[width=3.5cm,keepaspectratio]{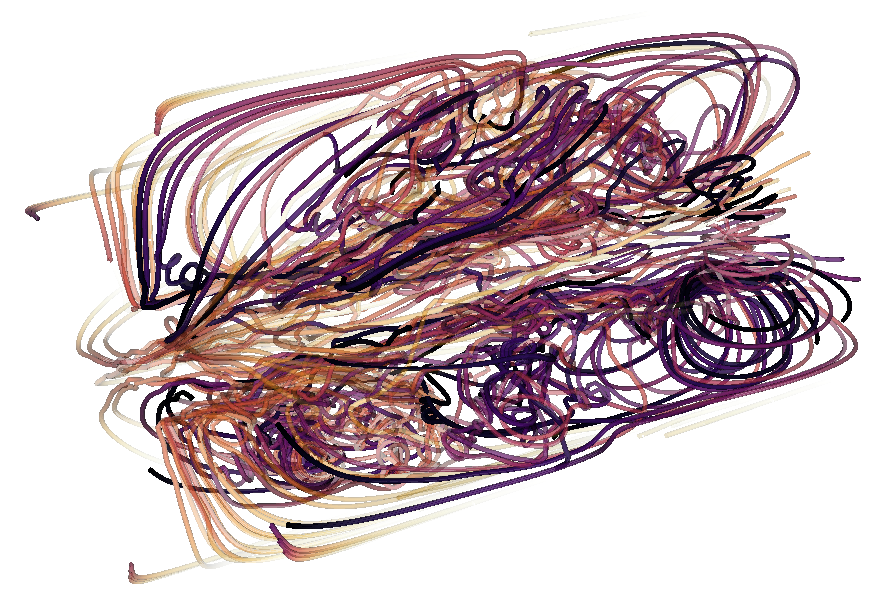}}
\subfloat[streamsurface]{\includegraphics[width=3.5cm,keepaspectratio]{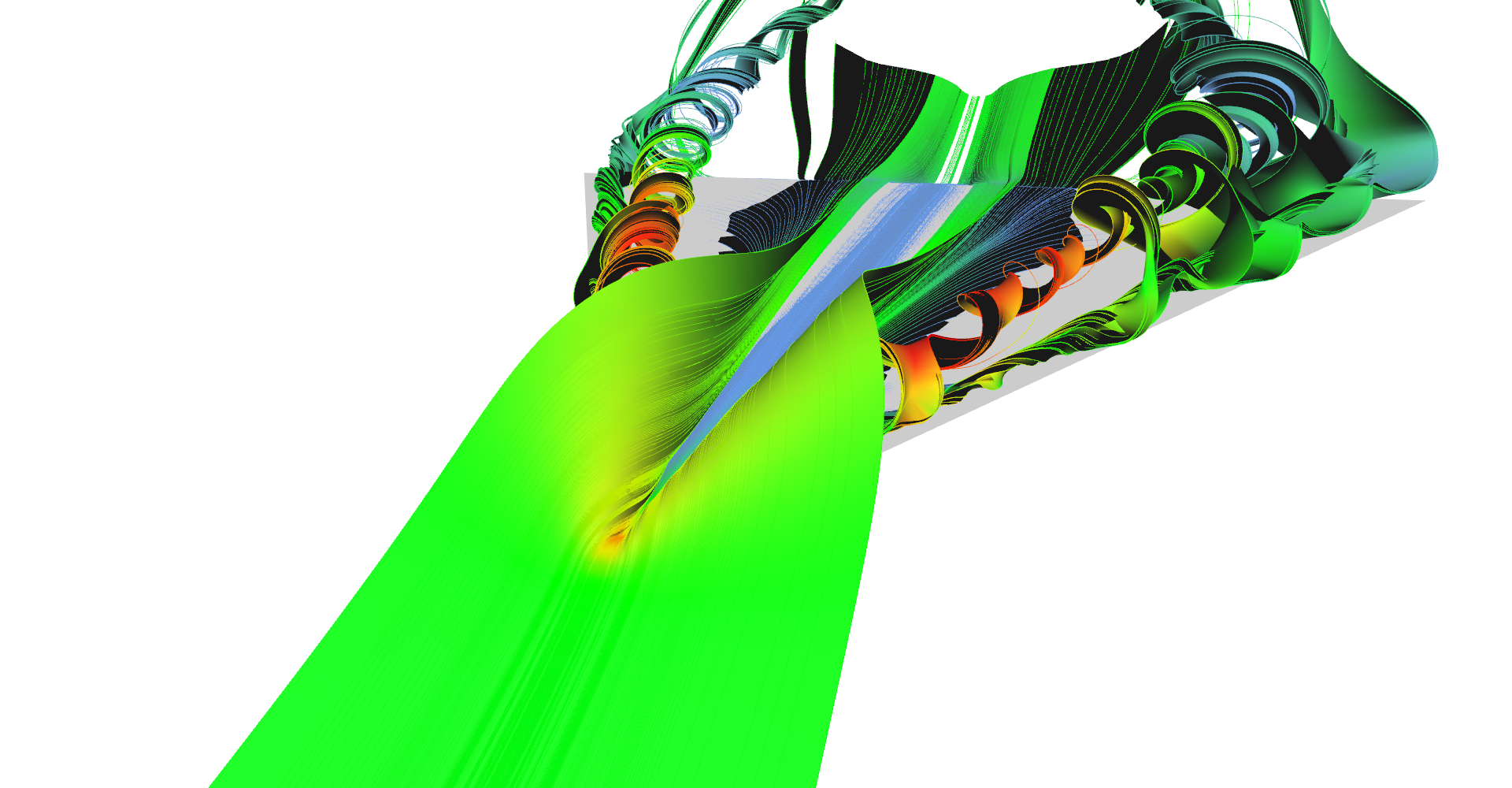}}
\subfloat[FTLE]{\includegraphics[width=4.5cm,keepaspectratio]{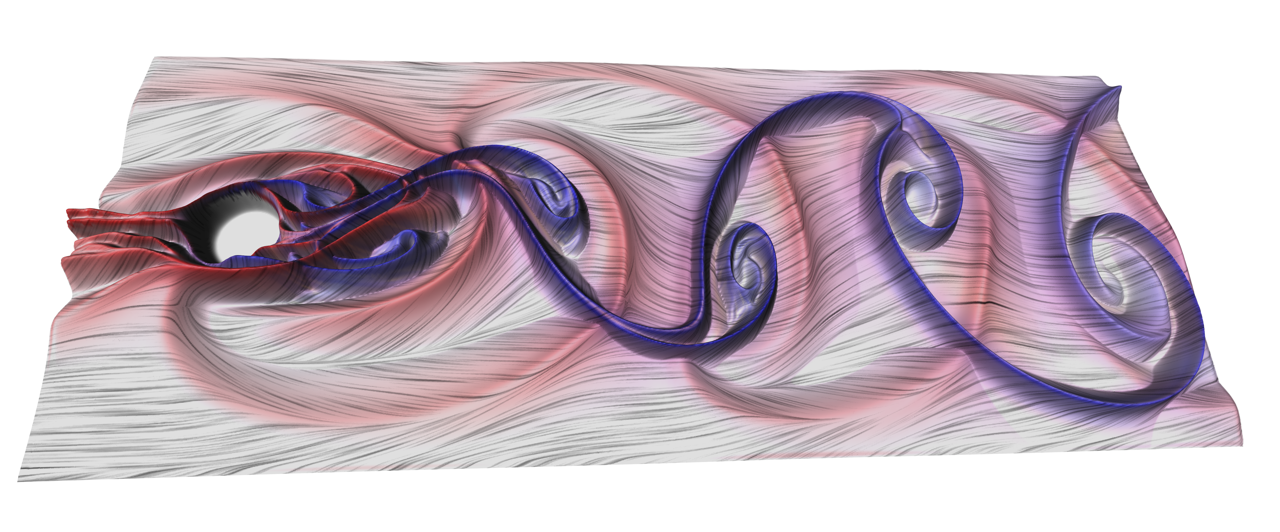}}
\subfloat[Poincar\'e]{\includegraphics[width=4.5cm,keepaspectratio]{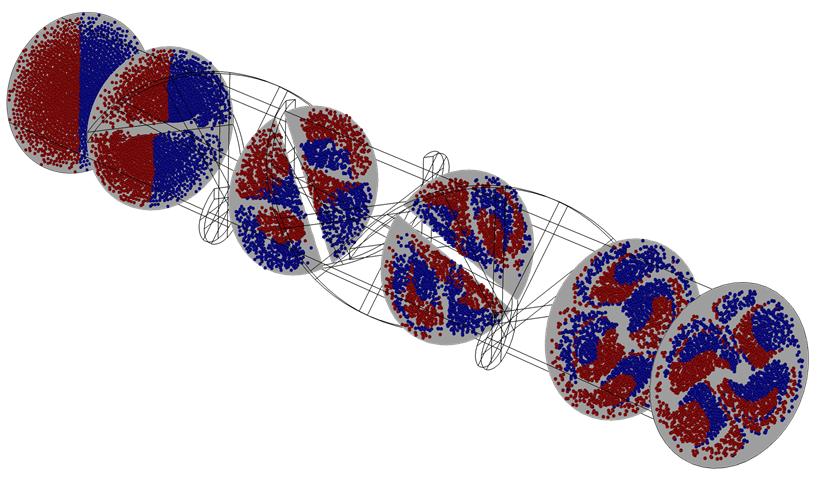}}
\caption{\label{fig:flowvis} Example flow visualizations from four representative algorithms.
Subfigure (a) shows streamlines rendered over a slice of jet plume data created using the Gerris Flow Solver~\cite{popinet2003gerris},
subfigure (b) shows a streamsurface which is split by turbulence and vortices that can be observed towards the end~\cite{fiser2013vis},
subfigure (c) shows attracting (blue) and repelling (red) Lagrangian structures extracted as FTLE ridges from a simulation
of a von Korman vortex street~\cite{kasten2010lagrangian}, and
subfigure (d) shows a Poincar\'e plot of a species being dissolved in water, where
the color of the dots represent the level of dissolution~\cite{carver2015vis}.}
\end{figure*}

Different flow visualization algorithms use these components in different ways,
resulting in different performance across the algorithms.
One way of comparing the different algorithms' perforamance can be the workload
required by the algorithms,
which roughly translates to the total computation required by the algorithm.
This workload can be defined as the total number of advection steps completed by
the algorithm, which is the product of the total number of particles required
by the algorithm and the number of steps expected to be completed by each particle.
Figure \ref{fig:flowvis} shows examples of four different flow visualization algorithms
that demonstrate significant differences in their workloads and behaviours.
Table \ref{tab:params} highlights the differences between the workloads for the example algorithms.
\begin{table}[h!]
\centering
\begin{tabular}{rlll}
\toprule
Seeding Strategy & \boit{Sparse} & \boit{Packed} & \boit{Seeding Curves}\\[1ex]
\midrule
Number of Seeds & \boit{Small}     & \boit{Medium} & \boit{Large} \\
                & $\leq$1/1K cells & \textasciitilde1/100 cells   & $\geq$1/cell \\[1ex]
\midrule
Number of Steps & \boit{Small} & \boit{Medium} & \boit{Large} \\
                & $\leq$100    & \textasciitilde1K & $\geq$10K \\[1ex]
\bottomrule
\end{tabular}
\\{(a) Each parameter is classified in three catagories.}
\begin{tabular}{rlll}
\toprule
Algorithm & Seeding & \# Seeds & \# Steps\\[1ex]
\midrule
Streamlines   & Sparse/Packed  & Small  & Large \\[1ex]
Streamsurface & Seeding Curves & Medium & Large \\[1ex]
FTLE          & Packed         & Large  & Small \\[1ex]
Poincar\`{e}  & Packed         & Medium & Large \\[1ex]
\bottomrule
\end{tabular}
\\{(b) Typical parameter configurations for different flow visualization algorithms.}
\caption{\label{tab:params}
Parameters for seeding strategy, the number of seeds, 
and the number of steps for four representative flow visualization algorithms.
(a) describes the parameters and their classifications, and
(b) presents the typical values for the four algorithms.}
\end{table}

\section{Cost Model for Particle Advection Performance}
\label{sec:costmodel}

This section considers costs from the perspective
of the building blocks used to carry out particle advection.
Its purpose is to build a general framework for
reasoning about costs and also to inform which aspects contribute most
to overall cost.
That said, the simplicity of the cost model precludes directly evaluating
many of the optimizations described in later sections (I/O, parallelism,
precomputation, and adaptive step sizing);
this topic is revisited in Section~\ref{sec:conclusion}.
Finally, Appendix~\ref{app:costmodel} goes into more depth on the cost model,
including estimating costs for each term in the model,
notional examples, and validation of the model.

Let the costs for particle advection be denoted by $Cost$.
Then a coarse formulation for $Cost$ is:
\begin{equation}
\label{eq:advance1}
Cost = \sum_{i=0}^{i=P} \sum_{j=0}^{j=N_i} advance_{i,j}
\end{equation}
where $P$ represents the total number of particles used for the flow visualization,
$N_i$ represents the total number of steps taken by the $i^{th}$ particle,
and $advance_{i,j}$ represents the amount of work required by particle $i$ at step $j$
in the process of advancing the particle.

To better illuminate the overall costs,
the remainder of this section
considers how the coarse formulation in Equation~\ref{eq:advance1}
can be further decomposed.
%
We first consider the tasks within $advance_{i,j}$.
In particular, each step that advances a particle contains three components ---
taking an advection step,
analyzing the step in a way specific to the individual flow visualization algorithm,
and checking if the particle should be terminated.
Hence, Equation \ref{eq:advance1} can be written as:
\begin{equation}
\label{eq:advance2}
Cost = \sum_{i=0}^{i=P} \sum_{j=0}^{j=N_i} \Big( step_{i,j} + analyze_{i,j} + term_{i,j} \Big)  \\
\end{equation}
where
$step_{i,j}$ is the cost for advecting,
$analyze_{i,j}$ is the cost for analyzing, and
$term_{i,j}$ is the cost for checking the termination criteria
for the $i^{th}$ particle at the $j^{th}$ step.

The cost can be further broken down by exploring the cost for a single advection step, $step_{i,j}$.
Particle advection uses an ODE solver to determine the next position of a particle, and this solver requires the velocity of the particle at the current location.
Further, depending on the ODE solver,
additional velocity evaluations in the proximity of the particle may be required.
An Euler solver requires only one velocity evaluation,
while an RK4 solver requires four velocity evaluations.
Generalizing, the cost of a single particle advection step can be written as:
\begin{equation}
\label{eq:step1}
step_{i,j} = solve_{i,j} + \sum_{k=0}^{k=K} eval_{i,j,k}
\end{equation}
where
$solve_{i,j}$ is the cost for the ODE solver to determine the next position,
$K$ is the number of velocity evaluations required by the ODE, and
$eval_{i,j,k}$ is the cost for velocity evaluation for the $i^{th}$ particle for the $j^{th}$ step at
the $k^{th}$ location.

The cost for velocity evaluations, $eval_{i,j,k}$, can be further broken down into two components.
Each evaluation involves two operations:
locating the current cell for the current evaluation, and
interpolating the velocity values for the current position using velocities at the
vertices of the current cell.
In all, the cost of velocity evaluations can be written as:
\begin{equation}
\label{eq:eval}
eval_{i,j,k} = locate_{i,j,k} + interp_{i,j,k}
\end{equation}
where $locate_{i,j,k}$ is the cost for locating the cell,
and $interp_{i,j,k}$ is the cost for interpolating the velocities at the $k^{th}$ location.

Further, we can substitute \ref{eq:eval} in \ref{eq:step1} to yield:
\begin{equation}
\label{eq:step2}
step_{i,j} = solve_{i,j} + \sum_{k=0}^{k=K} (locate_{i,j,k}+interp_{i,j,k})
\end{equation}

Finally, we can substitute \ref{eq:step2} in \ref{eq:advance2}
to obtain our final formulation:
\begin{equation}
\label{eq:complete}
\begin{split}
Cost = \sum_{i=0}^{i=P} \sum_{j=0}^{j=N_i} &
\Big( solve_{i,j} + \sum_{k=0}^{k=K} \big( locate_{i,j,k} + interp_{i,j,k} \big) \\
                                   & + analyze_{i,j} + term_{i,j} \Big)
\end{split}
\end{equation}

\section{Algorithmic Optimizations}
\label{sec:optimizations}

This section surveys algorithmic optimizations for particle advection building blocks,
i.e., techniques for executing a given workload using fewer operations.
Some of the building blocks do not particularly lend themselves to algorithmic optimizations.
For example, a RK4 solver requires a fixed number of FLOPS,
and the only possible ``optimization'' would be to use a different solver or adaptive step sizes.
That said, cell location allows room for possible optimizations.
Further,
the efficiency of vector field evaluation can be improved by considering underlying I/O operations.
This section discusses four optimizations that address the algorithmic challenges
Section \ref{adaptive} discusses optimizations to ODE solvers,
Section \ref{locators} discusses optiizations for cell location,
Section \ref{sec:ioeff} discusses strategies to improve I/O efficiency,
and finally Section \ref{precompute} discusses strategies that involve precomputation.

\subsection{ODE Solvers}
\label{adaptive}
The fundamental problem underlying particle advection is solving of ODEs.
Many methods are available for this, with different trade-offs, and a
comprehensive review is beyond the scope of this work, and we refer the reader
to the excellent book by Hairer et al.~\cite{hairer2000} for a more thorough
overview. 
Due to the generally (numerically) benign nature of vector fields used in
visualization, a set of standard schemes is used in many visualization
implementations.

Beyond the Euler and the fourth-order Runge-Kutta (RK4) methods, techniques with
adaptive step size control have proven useful. 
The primary objective of such methods is to allow precise control over the error
of the approximation of the solution, which is achieved by automated selection
of the step size (which in turn controls the approximation error) in each step.
Often used methods in this context are the Runge-Kutta Fehlberg (RFK)
method~\cite{hairer2000}, the Runge-Kutta Cash-Karp (RKCK) method~\cite{hairer2000}, and
the Dormand-Prince fifth-order scheme (DOPRI5)~\cite{prince1981high}.
As an additional benefit relevant in this context, due to the typically low
error tolerances required for visualization purposes, significant performance
benefits can be obtained if the adaptive step sizing results in fewer large
steps taken compared to fixed-step size methods.
While the magnitude of such benefits depends on a variety of factors that are
hard to quantify, using an adaptive step sizing method is generally recommended.
Corresponding implementations are widely available, e.g. in the VTK framework~\cite{geveci2012vtk, hanwell2015visualization}. 

For specialized applications, substantial performance benefits may be obtainable
by relying on domain-specific integration schemes that generally exhibit higher
accuracy orders and thus allow larger step sizes than general-purpose schemes. 
For example, Sanderson et al.~\cite{sanderson2010analysis} report substantial
speedup from employing an Adams-type scheme for visualizing high-order fusion
simulation data. However, general guidance on the selection of optimal schemes
for domain-specific vector field data remains elusive.

\subsection{Cell Locators}
\label{locators}

Cell locators facilitate interpolation queries over a grid and rely on auxiliary
data structures that partition candidate cells spatially.
These are typically constructed in a pre-processing step and induce linear
memory overhead in the number of cells $N$, while accelerating queries to
$\mathcal{O}(\log N)$.
Many cell location schemes allow trading off memory overhead for improved
performance.
A variety of schemes have been developed for different scenarios. For example,
limited available memory, e.g.
on GPUs, can be addressed through multi-level data structures.
According to Lohner and Ambrosiano~\cite{lohner1990vectorized} the process of
cell location can follow one of the following three approaches.

\boit{Using a Cartesian background grid:}
Cell are spatially subdivided using a superimposed Cartesian grid, storing a
list of overlapping cells of the original grid per superimposed cell.
The superimposed cell can be found in constant time, and cell location then
requires traversing all overlapping cells to find the actual containing cell for the
query point.
While conceptually simple, this approach is not ideal if the background grid
exhibits large variances in cell sizes, either incurring excessive storage
overhead or decreased performance, depending on the resolution of the
superimposed grid.

\boit{Using tree structures:}
A basic approach to hierarchical cell location is the use of octrees for cell
location~\cite{Schroeder:1997,Wilhelms:1992}.
Each leaf of an octree stores cells whose bounding box overlaps with the leaf
extents.
Leaves are subdivided until either a maximum depth is reached, or the number of
overlapping cells falls below an upper bound.
Cell location proceeds by traversing the octree from the root and descending
through nodes until a leaf is reached, which then contains all the candidate
cells.
Due to the regular nature of octree subdivision, this approach does not work
well with non-uniform vertex distributions, requiring either too many levels of
subdivision and thus a considerable memory overhead, or does not shrink the
candidate cell range down to acceptable levels.

Using kd-trees instead of octrees facilitates non-uniform subdivision, at the
cost of generally deeper trees and a storage overhead.
An innovative approach was given by Langbein et al.\cite{Langbein:2003}, based
on a kd-tree storing just the vertices of an unstructured grid.
This allows to quick location of a grid vertex close to the query point; using
cell adjacency, ray marching is used to traverse the grid towards the query
point using cell walking.
Through clever storage of the cell-vertex incidence information, storage
overhead can be kept reasonable.

Garth and Joy described the \emph{cell tree}~\cite{garth2010fast}, which employs
a kd-tree-like bounding interval hierarchy based on cell bounding boxes to
quickly identify candidate cells.
This allows a flexible trade-off between performance and storage overhead and
allows rapid cell location even for very large unstructured grids with hundreds
of millions of cells on commodity hardware and on memory-limited GPU
architectures.

Addressing storage overhead directly, Andrysco and Tricoche~\cite{Andrysco:2010}
presented an efficient storage scheme for kd-trees and octrees, based on
\emph{compressed sparse row} (CSR) storage of tree levels, termed \emph{Matrix
*Trees}.
The tree data structure is encoded as a sparse matrix in CSR representation.
This alleviates most of the memory overhead of kd-trees, and they are able to
perform cell location with reduced time and space complexity when compared with
typical tree data structures.

Overall, non-uniform hierarchical subdivision can accommodates large meshes  
with significant variations in cell shapes and sizes well.
While Lohner and Ambrosiano note that vectorization of this approach is
challenging as tree-based schemes introduce additional indirect addressing,
vectorization is still possible on modern CPU and GPU architectures with good
performance~\cite{garth2010fast}.

\boit{Using successive neighbor searches:}
For the case of particle integration, successive interpolation queries exhibit
strong  coherence and are typically spatially close.
This enables a form of locality caching: For each interpolation query except the
first, the cell that contained the previous query point is checked first.
If it does not contain the interpolation point, its immediate neighbors are
likely to contain it, potentially reducing the number of cells to check.
The initial interpolation point can be located using a separate scheme, e.g.
as discussed above.

Lohner and Ambrosiano, as well as Ueng et al.~\cite{ueng1996efficient}, adopted
a corresponding successive neighbor search method to cell location in particle
advection for efficient streamline, streamribbon, and streamtube construction.
They restricted their work to linear tetrahedral cells for simplification of
certain formulations, requiring a pre-decomposition for general unstructured
grids.
Note that when applied to tetrahedral meshes, the successive neighbor search
approach is sometimes also referred to as \boit{tetrahedral
walk}~\cite{schirski2006interactive,bussler2011interactive}.

Kenwright and Lane~\cite{kenwright1996interactive} extended the work by Ueng et
al. by improving the technique to identify the particle's containing tetrahedron.
Their approach uses fewer floating point operations for cell location compared to Ueng et al.

Successive neighbor search is also naturally incorporated in the method of
Langbein et al.~\cite{Langbein:2003}; ray casting with adjacency walking towards
begins at the previous interpolation point in this case.

\begin{table*}
\centering
\begin{tabular}{lllrrrr}
\toprule
Algorithm & Application &  Intent /   & Data & Time  & Seed   & Performance \\
          &             &  Evaluation & Size & Steps & Count  &             \\
\midrule
\multirow{3}{4cm}{Lohner and Ambrosiano\cite{lohner1990vectorized}} &
\multirow{3}{*}{Streamlines} & 
\multirow{3}{5cm}{Fast cell location and efficient vectorization} 
  & $870^*$ & - & 10K & $14\times$  \\
&&&&&&\\
&&&&&&\\[1ex]

\multirow{3}{4cm}{Ueng et al.\cite{ueng1996efficient}} &
\multirow{3}{*}{Streamlines} & 
\multirow{3}{5cm}{Streamline computation and cell location in canonical coordinate space} 
  & $320K^*$ & - & \multirow{3}{*}{100} & $1.61\times$  \\
&&& $225K^*$ & - &                      & $1.59\times$  \\
&&& $288K^*$ & - &                      & $1.58\times$  \\[1ex]

\multirow{3}{4cm}{Chen et al.\cite{chen2011flow}} &
\multirow{3}{*}{Streamlines} & 
\multirow{3}{5cm}{Improving data layout for better I/O performance} 
  & 134M & - & \multirow{3}{*}{4K} & $0.96 - 1.30\times$  \\
&&& 200M & - &                     & $0.98 - 1.98\times$  \\
&&& 537M & - &                     & $0.99 - 1.29\times$ \\[1ex]

\multirow{3}{4cm}{Chen et al.\cite{chen2012flow}} &
\multirow{3}{*}{Pathlines} & 
\multirow{3}{5cm}{Improving data layout for better I/O performance} 
&  25M & 48 & \multirow{3}{*}{4K}   & $1.25 - 1.38\times$  \\
&&&  65M & 29 &                     & $1.10 - 1.31\times$  \\
&&&  80M & 25 &                     & $1.19 - 1.36\times$  \\[1ex]

\multirow{3}{4cm}{Chen et al.\cite{chen2013graph}} &
\multirow{3}{*}{FTLE} & 
\multirow{3}{5cm}{Improving data layout for better I/O performance} 
&    25M & 48 & \multirow{3}{*}{-} & \multirow{3}{*}{$1.08-1.32\times$} \\
&&&  65M & 29 &                           &   \\
&&&  80M & 25 &                           &   \\[1ex]
\bottomrule
\end{tabular}
\caption{
Summary of studies considering algorithmic optimizations to particle advection.
Studies that do not report quantitative performance improvements are not mentioned in the table.
The asterisk for entries in the data size column represent unstructured grids.
\label{tab:algosummary}
}
\end{table*} 

\subsection{I/O Efficiency}
\label{sec:ioeff}
Simulations with very large numbers of cells often output
their vector fields in a block-decomposed fashion,
such that each block is small enough to fit in the memory of a compute node.
%
Flow visualization algorithms that process block-decomposed data vary in strategy,
although many operate by storing a few of these blocks in memory at a time,
and loading/purging blocks as necessary.
%
%
This method of computation is known as out-of-core computation.
One of the significant bottlenecks for flow visualization algorithms
while performing out-of-core computations is the cost of I/O.
Particle advection is a data-dependent operation and efficient prefetching to
ensure sequential access to data can be very beneficial in minimizing these I/O costs.
%
This section discusses the works that aim to improve particle advection
performance by improving the the efficiency of I/O operations.

Chen et al.~\cite{chen2011flow} presented an approach to improve the
I/O efficiency of particle advection for out-of-core computation.
Their approach relies on constructing an access dependency graph (ADG)
based on the flow data.
The graph’s nodes represent the data blocks,
and the edges are weighted based on the probability that a particle travels from one block to another.
The information from the graph is used during runtime to minimize data block misses.
Their method demonstrated speed-ups over the Hilbert curve layout~\cite{hilbert1891stetige}.
%

Chen et al.~\cite{chen2012flow} extended the previous work to out-of-core computation of pathlines. 
%
%
Their results show a performance improvement
in the range of 10\%-40\% compared to the Z-curve layout~\cite{zhang1994z, zhang2003z}.

Chen et al.\cite{chen2013graph} expanded the work further to introduce
a seed scheduling strategy to be used along with the graph-based data layout.
They demonstrated an efficient out-of-core approach for calculating FTLE.
The performance improvements observed against the Z-curve layout were in the range of 8\%-32\%.

\subsection{Precomputation}
\label{precompute}
Besides optimizing individual particle advection building blocks, optimization of certain flow visualization workloads can benefit from a two-stage approach.
During the first stage, based on current literature, a set of particle trajectories can be computed to inform data access patterns, or serve as a basis for interpolating new trajectories.
Depending on the objectives, the number of trajectories computed during the first stage varies. 
The resulting set of trajectories can be referred to as the precomputed trajectories.
%

%
Precomputed trajectories can inform data access patterns to provide a strategy to improve I/O efficiency, as mentioned in the context of the study by Chen et al.~\cite{chen2011flow} in the previous section.
A similar approach was studied by Nouansengsy et al.~\cite{nouanesengsy2011load} to improve load balancing in a distributed memory setting.
In these cases, the first stage is a preprocessing step and a small number of particle might be advected to form the set of precomputed trajectories.
%
%

For a computationally expensive particle advection workload, a strategy to accelerate the computation or improve interactivity of time-varying vector field visualization is to divide the workload into two sets.
The first set includes particle trajectories computed using high-order numerical integration. 
%
%
%
The second set includes particle trajectories that are derived by interpolating the precomputed trajectories. 
If new particle trajectories can be derived from the precomputed set faster than numerical integration, while remaining accurate and satisfying particle trajectory requirements for the specific flow visualization use case, then the total computational cost of the workload can be reduced compared to the numerical intergration of every trajectory.

Hlawatsch et al.~\cite{hlawatsch2010hierarchical} introduced a hierarchical scheme to construct integral curves, streamlines or pathlines, using sets of precomputed short flow maps.
They demonstrated the approach for the computation of the finite-time Lyapunov exponent and the line integral convolution.
Although the method introduces a trade-off of reduced accuracy, they demonstrate their approach can result in an order of magnitude speed up for long integration times.

To accelerate the computation of streamline workloads,
Bleile et al.~\cite{bleile2017accelerating} employed block exterior flow maps (BEFMs) produced using precomputed trajectories.
BEFMs, i.e., a mapping of block-specific particle entry to exit locations, are generated to map the transport of particles across entire blocks in a single interpolation step.
Thus, when a new particle enters a block, instead of performing an unknown number of numerical integration steps to traverse the region within the block, based on the mapping information provided by precomputed trajectories, the location of the particle exiting (or terminating within) the block can be directly interpolated as a single step.
Depending on the nature of the workload, large speedups can be observed using this strategy.
For example, Bleile et al.~\cite{bleile2017accelerating} observed up to 20X speed up for a small loss of accuracy due to interpolation error.
%

To support exploratory visualization of time-varying vector fields, 
Agranovsky et al.~\cite{agranovsky2014improved} proposed usage of in situ processing to extract accurate Lagrangian representations. 
In the context of large-scale vector field data, and subsequent temporally sparse settings during post hoc analysis, reduced Lagrangian representations offer improved accuracy-storage propositions compared to traditional Eulerian approaches, as well as, can support acceleration of trajectory computation during post hoc analysis.
By seeding the precomputed trajectories along a uniform grid, structured (uniform or rectilinear) grid interpolation performance can be achieved during post hoc analysis.
To further optimize the accuracy of reconstructed pathlines in settings of temporal sparsity, research has considered how varying the set of precomputed trajectories can improve accuracy-storage propositions. 
%
%
%
For example, Sane et al.~\cite{sane2019interpolation} studied the use of longer trajectories to reduce error propagation and improve accuracy, and Rapp et al.~\cite{rapp2019void} proposed a statistical sampling technique to determine where seeds should be placed.
Unstructured sampling strategies, however, can increase the cost of post hoc interpolation and diminish computational performance benefits.

\subsection{Summary}
Table \ref{tab:algosummary} summarizes studies that address algorithmic optimizations and report performance improvements against a baseline implementation.
The studies mentioned in the table either target optimizations for cell location or perform better I/O operations.
For ODE solvers and precomputation, reporting performance improvements is difficult because of an associated accuracy trade-off for better performance.
Optimizations to cell locators for unstructured grid enable significant speed-ups for the workloads.
With a combination of efficient cell location and vectorization, Lohner and Ambrosiano~\cite{lohner1990vectorized} achieved the speed of $14\times$.
However, the other study~\cite{ueng1996efficient} demonstrated a speed-up of around $1.6\times$.
The works by Chen et al.~\cite{chen2011flow, chen2012flow, chen2013graph} for efficient I/O for particle advection all demonstrated speed-ups up to $1.3\times$.

\section{Using Hardware Efficiently}
\label{sec:usinghwefficiently}
Flow visualization algorithms often share resources with large simulation codes,
or require large amounts of computational resources of their own depending on the
needs of the analysis task.
This means flow visualization algorithms are often required to execute on supercomputers.
Executing codes on supercomputers is expensive and it is necessary that
all analysis and visualization tasks execute with utmost efficiency.
Modern supercomputers have multiple ways to make algorithms execute fast.
Typically,
supercomputers have thousands of nodes over which computation can be distributed, 
and each node has multi-core CPUs on along with multiple accelerators (e.g., GPUs) for parallelization.
As a result, algorithms are expected to make efficient use of this billion-way concurrency.
This section discusses research for particle advection that addresses efficient
usage of available hardware.
Section \ref{sec:shared} discusses research that aims to improve shared-memory (on-node) parallelism.
Section \ref{sec:distmemparadv} discusses research that aims to improve distributed memory parallelism.
Section \ref{sec:hybrid} discusses research that uses both shared and distributed memory parallelism.

\subsection{Shared Memory Parallelism for Particle Advection}
\label{sec:shared}

%
Shared memory parallelism refers to using parallel resources on a single node.
The devices that enable shared memory parallelism are multi- and many-core CPUs and other
accelerators, such as GPUs.
In the case of shared memory parallelism,
multiple threads of a program running on different cores of a processor (CPU or a GPU) share memory,
hence the nomenclature.
One of the primary reasons for the increase in supercomputers'
compute power can be attributed to the advancements of CPUs and accelerator hardware.
%
%
In all, for applications to make cost-effective use of resources,
it has become exceedingly important to use shared memory resources efficiently.
However, making efficient use creates many challenges for the programmers and users.
Two important factors to consider are
1) efficient use of shared memory concurrency, and 2) performance portability.
\begin{table*}
	\centering
	\begin{tabular}{lllrrrr}
		\toprule
		Algorithm                      & Application &  Intent /      & Data & Time  & Seed  & Performance\\
		&             &  Evaluation    & Size & Steps & Count &            \\
		\midrule
		\multirow{2}{2.5cm}{Kr{\"u}ger et al. \cite{kruger2005particle}}  &
		\multirow{2}{2cm}{source-dest} &
		\multirow{2}{5cm}{Interactive flow visualization (steady) using GPUs} &
		- & - & - &  60-80$\times$\\[1ex]
		&&&&&& \\[1ex]

		\multirow{2}{2.5cm}{B{\"u}rger et al.\cite{burger2007interactive}} &
		\multirow{2}{2cm}{various}     &
		\multirow{2}{5cm}{Interactive flow visualization (unsteady)}       &
		&  &   &  \\
		&&&&&&\\[1ex]

		\multirow{2}{2.5cm}{B{\"u}rger et al.\cite{burger2008importance}} &
		\multirow{2}{2cm}{various}     &
		\multirow{2}{5cm}{Interactive flow visualization using importance metrics}    &
		7M    & -  &    &   \\
		&&& 4M    & 22 &    &   \\
		&&& 1M    & 30 &    &   \\[1ex]

		\multirow{2}{2.5cm}{B{\"u}rger et al.\cite{buerger2009interactive}} &
		\multirow{2}{2cm}{streak surface}   &
		\multirow{2}{5cm}{Interactive streak surface visualization} &
		589K & 102   & \multirow{2}{*}{400} &  \\
		&&& 4.1M & 22    &  &  \\[1ex]

		\multirow{3}{2.5cm}{Schirski et al.\cite{schirski2006interactive}} &
		\multirow{3}{2cm}{pathlines, source-dest} &
		\multirow{3}{5cm}{Efficient cell location on GPUs} &
		$0.8M^*$     & 5      & \multirow{3}{*}{1M}  &              \\
		&&& $1.1M^*$     & 101    &   &              \\
		&&& $3.7M^*$     & 200    &   &              \\[1ex]

		\multirow{2}{2.5cm}{Garth et al.\cite{garth2010fast}} &
		\multirow{2}{2cm}{source-dest} &
		\multirow{3}{5cm}{Efficient cell location on GPUs for unstructured grids / Comparison against CPUs} &
		\multirow{2}{*}{$23.6M^*$}   & \multirow{2}{*}{-}   & 250K  & \multirow{2}{*}{$16.5\times$} \\
		&&&    &                      & 1M    &  \\
		&&&&&&\\[1ex]

		\multirow{3}{2.5cm}{Bu{\ss}ler et al.\cite{bussler2011interactive}} &
		\multirow{3}{2cm}{source-dest} &
		\multirow{3}{5cm}{Efficient cell location on GPUs using improved tetrahedral walk} &
		$4.2M^*$ & 5   & \multirow{3}{*}{1M}      &  \\
		&&& $115M^*$ & 101 &                          &  \\
		&&& $743M^*$ & 200 &                          &  \\[1ex]

		\multirow{6}{2.5cm}{Pugmire et al.\cite{pugmire2018portable}} &
		\multirow{6}{2cm}{source-dest} &
		\multirow{6}{5cm}{Performance Portability / Comparison with specialized comparators for CPUs and GPUs} &
		\multirow{2}{*}{134M} & \multirow{6}{*}{-} & \multirow{6}{*}{10M} & $0.37-0.48\times$ (GPUs) \\
		&&&&&& $0.29-0.36\times$ (CPUs) \\
		&&& \multirow{2}{*}{134M} &   &   & $1.56-2.24\times$ (GPUs)\\
		&&&&&& $0.79-0.84\times$ (CPUs) \\
		&&& \multirow{2}{*}{134M} &   &   & $1.42-2.04\times$ (GPUs) \\
		&&&&&& $0.51-0.59\times$ (CPUs) \\[1ex]

		\bottomrule
	\end{tabular}
	\caption{
		Summary of shared memory particle advection.
		The asterisk for entries in the data size column represent unstructured grids.
		\label{tab:sharedsum}
	}
\end{table*}

\begin{figure*}[htb]
	\centering
	\includegraphics[width=0.75\linewidth]{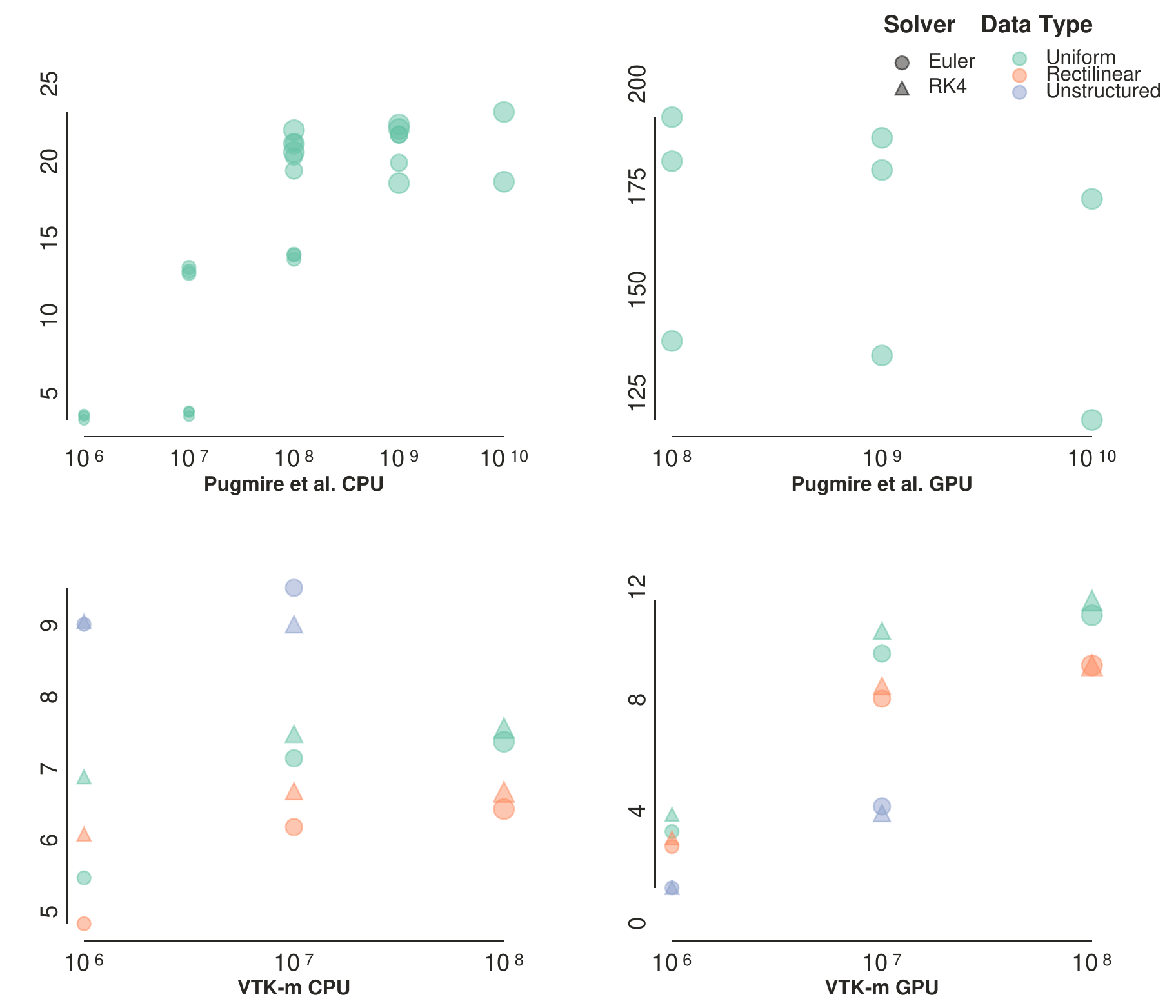}
	\caption{
		Scatter plots for CPU and GPU speed-ups for particle advections workloads.
		The X axis represents the magnitude of the workload in terms of total number of
		steps for each of the sub plots and the Y axis represents the speed-up.
		The size of the glyphs corresponds to the number of particles used in the experiment.
		The data for the plots on the top is collected from numbers reported by
		Pugmire et al.\cite{pugmire2018portable}
		They used 28 CPU cores and a Nvidia P100 GPU.
		The data from the plots on the bottom is collected from new experiments using VTK-m.
		This study used 12 CPU cores and a Nvidia K80 GPU.
		\label{fig:sharedsum}
	}
\end{figure*}

GPUs have become a popular accelerator choice in the past decade,
with most leading supercomputers using GPUs as accelerators~\cite{top500}.
Part of this has been the availability of specialized toolkits,
including early efforts like Brook-GPU~\cite{buck2004brook} and
popular efforts like Nvidia's CUDA~\cite{nickolls2007gpu},
that enable GPUs to be used as general purpose computing devices~\cite{bergman2008exascale}.
However,
programming applications for efficient execution on a GPU remains challenging for three main reasons.
First, unlike CPUs which are built for low latency,  GPUs are built for high throughput.
CPUs have fewer than a hundred cores, while GPUs have a few thousand.
However,
each CPU core is significantly more powerful than a single GPU core.
Second, efficient use of the GPU requires applications to have sufficiently large parallel workloads.
Third, executing a workload on a GPU also has an implicit
cost of data movement between the host and the device,
where a host is the CPU and the DRAM of the system,
and the device is the GPU and its dedicated memory.
This cost makes GPUs inefficient for smaller workloads.

This sub-section discusses particle advection using shared memory parallelism in two parts.
Section~\ref{sec:gpupa} discuss works to optimize the performance particle advection on GPUs and Section~\ref{sec:gpucl} describes optimizations for cell locators.
Section \ref{sec:cpupa} discusses works that use CPUs for improving the
performance of particle advection.

\subsubsection{Shared memory GPUs}
\label{sec:gpupa}
Most of the solutions that focused on shared memory optimization focused on improving the performance on GPUs.
This is because particle advection can benefit from using GPUs when there are many particles to advect.
As particles can be advected independently from one another,
each particle can be scheduled with a separate thread of the GPU,
making the most of the available concurrency.
Many works have tried to address performance issues of particle advection
using GPUs, however, with different goals.

Kr{\"u}ger et al.~\cite{kruger2005particle}
presented an approach for interactive
visualization of particles in a steady flow field using a GPU.
They exploited the GPU's ability to simultaneously perform advection and render
results without moving the data between the CPU and the CPU.
%
This was done by accessing the texture maps in the GPU's vertex units
and writing the advection result.
Their approach on the GPU provided the interactive rendering at 41 fps
(frames per second) compared to 0.5 fps on the CPU.

B{\"u}rger et al.~\cite{burger2007interactive}
extended the particle advection framework
described by Kr{\"u}ger et al. for unsteady flow fields.
With their method, unsteady data is streamed to the GPU using a ring-buffer.
%
While the particles are being advected in some time interval
$[t_{i}, t_{i+1}]$,
another host thread is responsible for moving
$t_{i+2}$
from host memory to device memory.
At any time, up to three timesteps of data are stored on the device.
%
By decoupling the visualization and data management tasks,
particle advection and visualization can occur without delays due to data loading.
%
B{\"u}rger et al.~\cite{burger2008importance} further demonstrated the efficacy
of their particle tracing framework for visualizing an array of flow features.
These features were gathered using some metric of importance,
e.g., FTLE, vorticity, helicity, etc.

B{\"u}rger et al.~\cite{buerger2009interactive}
also provided a way for interactively rendering streak surfaces.
Using GPUs, the streak surfaces can be adaptively refined/coarsened
while still maintaining interactivity.

\paragraph{Cell Locators for GPUs}
\label{sec:gpucl}
Bu{\ss}ler et al.~\cite{bussler2011interactive}
presented a GPU-based tetrahedral walk for particle advection.
Their approach for cell location borrowed heavily from the work by
Schiriski et al.~\cite{schirski2006interactive} discussed in Section \ref{locators}.
However, they could execute the cell location strategy entirely on the GPU
and do not require the CPU for the initial search.
%
Additionally,
they evaluated different Kd-tree traversal strategies to evaluate the
impact of these strategies on the tetrahedral walk
%
%
Their results concluded that the \boit{single-pass} method,
which performs only one pass through the kd-tree to find the nearest cell vertex
(without the guarantee of it being the nearest) performs the best.
The other strategies evaluated in the study were
\boit{random restart} and \boit{backtracking}.

Garth and Joy~\cite{garth2010fast} presented an approach for cell location
based on bounding interval hierarchies.
Their search structure, called \boit{celltree}, improves construction times via
a heuristic to determine good spatial partitions.
The authors presented a use case of advecting a million particles on a GPU in an
unstructured grid with roughly 23 million hexahedral elements.
The celltree data structure was able to obtain good performance on GPUs despite
no GPU-specific optimizations.

\subsubsection{Shared Memory Parallelism on CPUs}
\label{sec:cpupa}
Hentschel et al.~\cite{Hentschel2015} presented a solution that focused on optimizing particle advection on CPUs.
Their solution studied the performance benefits of using SIMD extensions on CPUs to acheive better performance.
This paper addresses the general tendency of particles to move around in the flow field.
This decreases memory locality of that data that are required to perform the advection computation.
This work demonstrated the advantage of packaging particles into spatiaily local groups where SIMD extensions are able to be more efficient.
Their approach resulted in performance improvements of up to 5.6X over the baseline implementation.

Finally, Pugmire et al.~\cite{pugmire2018portable} provided a platform portable solution
for particle advection using the VTK-m library.
The solution builds on data parallel primitives provided by VTK-m.
Their results demonstrated very good platform portability,
providing comparable performance to platform specific solutions on many-core CPUs and
Nvidia GPUs.

\subsubsection{Summary}

In terms of published research,
Table \ref{tab:sharedsum} presents a summary of shared memory particle advection.
These studies either presented approaches for interactive flow visualization
or optimizations for particle advection of GPUs using cell locators,
with one exception that demonstrated platform portability.

Figure \ref{fig:sharedsum}
shows a preliminary result for understanding the characteristics of shared memory parallel particle advection.
Some of the key observations are:
\begin{enumerate}
	\item
	Workloads with more number of particles scale better with added parallelism for the same amount of total work.
	\item
	The studies that used the RK4 integrator generally scaled better than the ones that used the Euler integrator.
	\item
	The experiments with unstructured data scaled better on the CPU than on the GPU.
	This could be because of the nature of memory accesses required by cell locators
	and justifies more research into GPU based cell locators.
\end{enumerate}

Additionally, the plots for the CPUs demonstrate consistency in terms of scalability
when the workload is increased.
The plot for the P100 GPUs (top right) suggests that it is not able to
scale larger workloads with the same efficiency as the smaller workloads
considered by Pugmire et al.~\cite{pugmire2018portable}
There is also a tremendous variation in the speed-ups achieved by two considered GPUs,
where the P100 GPU is able to achieve speed-ups of over $125\times$
and the K80 GPU achieves speed-ups of less than $12\times$.
The performance difference of particle advection between two generations of
GPUs can be significant.
Existing studies fail to capture this relation and makes it harder to estimate
to speed-up that can be realized.
Understanding the performance characteristics of particle advection across
different GPUs is planned as future work.

\subsection{Distributed Memory Parallelism for Particle Advection}
\label{sec:distmemparadv}
Fluid simulations are capable of producing large volumes of data.
Analyzing and visualizing volumes of data to extract useful information
demands resources equivalent to that of the simulation.
In most cases,
this means access to many nodes of a supercomputer to handle the
computational and memory needs of the analysis.
Particle advection based flow visualization algorithms often execute
in a distributed memory setting.
The objective of the distribution of work is to perform efficient computation,
memory and I/O operations, and communication.
There are multiple strategies for distributing particle advection workloads
in a distributed memory setting to achieve these objectives.
These can be categorized under two main classes:

\noindent \boit{Parallelize over particles:}
Particles are distributed among parallel processes.
Each process only advances particles assigned to it.
Data is typically loaded as required for each of the particles.

\noindent \boit{Parallelize over data:}
Blocks of partitioned data are distributed among parallel processes.
Each process only advances particles that occur within the data blocks assigned to it.
Particles are communicated between processes based on their data requirement.

Most distributed particle advection solutions are either an optimization
of these two classes or a combination of them.
The decision to choose between these two classes depends on multiple factors,
of which Camp et al.~\cite{camp2010streamline} identify the most prominent to be:

\noindent \boit{The volume of data set:}
If the data set can fit in memory,
it can be easily replicated across nodes and particles can be distributed among nodes,
i.e., the work can be parallelized over particles.
However, for large partitioned data sets,
work parallelized over data can be more efficient.

\noindent \boit{The number of particles:}
Some flow visualization algorithms require small number of particles integrated over a long duration,
while others require a large number of particles advanced for a short duration.
In the case where fewer particles are needed,
parallelization over data is a better approach as it could potentially reduce I/O costs.
In the case where more particles are needed,
parallelization over particles can help better distribute computational costs.

\noindent \boit{Distribution of particles:}
The placement of particles for advection can potentially cause performance problems.
When using parallelization over data,
if particles are concentrated within a small region of the data set,
the processes owning the associated data blocks will be responsible for a lot of
computation while most other processes remain idle.
Parallelization over particles can lead to better work distribution in such cases.

\noindent \boit{Data set complexity:}
The characteristics of the vector field have a significant influence on the work
for the processes, e.g., if a process owning a data block that contains a sink,
most particles will advect towards it, causing the process to do more work than the others.
In such a case, parallelize over particles will enable better load balance.
On the other hand,
when particles switch data blocks often (e.g., a circular vector field),
parallelize over data is better since it reduces the costs of I/O to load required blocks.

This section describes distributed particle advection works in two parts.
Section \ref{sec:distpar} describes the optimization for parallelizing
distributed particle advection in more depth.
Section \ref{sec:distsummary} summarizes findings from the survey of distributed
particle advection studies.
%
   
\subsubsection{Parallelization Methods}
\label{sec:distpar}
This section presents distributed particle advection works in three parts.
Section \ref{sec:overdata} presents works that optimize parallelization over data.
Section \ref{sec:overparticles} presents works that optimize parallelization over particles.
Section \ref{sec:hybridadv} presents works that use a combination
of parallelization over data and particles.
\paragraph{\boit{Parallelization over data}}
\label{sec:overdata}

\textit{``Parallelize over data''} is a paradigm for work distribution in flow
visualization where $M$ data blocks are distributed among $N$ processors.
Each process is responsible for performing computations for active particles
within the data blocks assigned to them.
This method aims to reduce the cost of I/O operations,
which is more expensive than the cost of performing computations.

Sujudi and Haimes~\cite{sujudi1996integration}
elicited the problems introduced by decomposing data into smaller blocks that can
be used within the working memory of a single node.
They presented important work in generating streamlines in a distributed memory
setting using the parallelize over data scheme.
They used a typical client-server model where clients perform the work,
and the server coordinates the work.
Clients are responsible for the computation of streamlines within their sub-domain;
if a particle hits the boundary of the sub-domain,
it requests the server to transfer the streamline to the process that owns the next sub-domain.
The server is responsible for keeping track of client request and sending streamlines
across to the clients with the correct sub-domain.
No details of the method used to decompose the data in sub-domains are provided.

Camp et al.~\cite{camp2010streamline} compared the MPI-only implementation to the
MPI-hybrid implementation of parallelizing over data.
They noticed that the MPI-hybrid version benefits from reduced communication of
streamlines across processes and increased throughput when using multiple cores
to advance streamlines within data blocks.
Their results demonstrated performance improvements between 1.5x-6x in the overall times
for the MPI-hybrid version over the MPI-only version.
The parallelize over data scheme is sensitive to the distribution of particles
and complexity of vector field.
The presence of critical points in certain blocks of data can potentially
lead to load imbalances.
Several techniques have been developed to deal with such cases and can be classified into two
categories
1) works that require knowledge of vector field, and
2) works that do not require knowledge of vector field. 

\noindent \boit{Knowledge of vector field required}
The works classified in this category acquire knowledge of vector
fields by performing a pre-processing step. Pre-processing allows
for either data or particles to be distributed such that all processes perform the same amount of computation.

Chen et al. presented a method that employs repartitioning
of the data based on flow direction, flow features, and the number of
particles~\cite{chen2008optimizing}.
They performed pre-processing of the vector field using various statistical and topological methods to enable effective partitioning.
The objective of their work is to produce partitions such that the streamlines produced would seldom have to travel between different data blocks. This enabled them to speed up the computation of streamlines due to the reduced communication between processes.

Yu et al.~\cite{yu2007parallel}
presented another method that relies on pre-processing the vector field.
They treated their spatiotemporal data as 4D data instead of considering the space 
and time dimensions as separate.
They performed adaptive refinement of the 4D data using a higher resolution for
regions with flow features and a lower resolution for others.
Later, cells in this adaptive grid were clustered hierarchically using a binary
cluster tree based on the similarity of cells in a neighborhood.
This hierarchical clustering helped them to partition data that ensure workload balance.
It also enabled them to render pathlines at different levels of abstraction.

Nouanesengsy et al.~\cite{nouanesengsy2011load} used pre-processing to estimate
the workload for each data block by advecting the initial set of particles.
The estimates calculated from this step are used to
distribute the work among processes.
Their proposed solution maintained load balance and improved performance.
While the solutions in this category are better at load balancing,
they introduce an additional step of pre-processing which has its costs.
This cost may be expensive and undesirable if the volume of data is significant.

\noindent \boit{Knowledge of vector field not required}
The works classified in this category aim to balance load dynamically
without any pre-processing.

Peterka et al.~\cite{peterka2011study} performed a study to analyze the effects of data
partitioning on the performance of particle tracing.
Their study compared static round-robin (also known as block-cyclic) partitioning
to dynamic geometric repartitioning.
The study concluded that while static round-robin assignment provided good load
balancing for random dense distribution of particles,
it fails to provide load balancing when data blocks contain critical points.
They also noticed that dynamic repartitioning based on workload could improve 
the execution time between 5\% to 25\%.
However, the costs to perform the repartitioning are restrictive.
They suggest more research needs to focus on using less synchronous communication
and improvements in computational load balancing.

Nouanesengsy et al.~\cite{nouanesengsy2012parallel} extended the work by Perterka et al.
to develop a solution for calculating Finite-Time Lyapunov Exponents (FTLE) for
large time-varying data.
The major cost in performing FTLE calculations is incurred due to particle tracing.
Along with \textit{parallelize over data}, they also used \textit{parallelize over time},
which enabled them to create a pipeline that
could advect particles in multiple time intervals in parallel.
Although their work did not focus on load-balancing among processes,
it presented a novel way to optimize time-varying particle tracing.
Their work solidifies the conclusions about static data partitioning of the study by Peterka et aL~\cite{peterka2011study}.

Zhang et al.~\cite{zhang2017dynamic} proposed a method that is better at
achieving dynamic load balancing.
Their approach used a new method for domain decomposition,
which they term as the constrained K-d tree.
Initially, they decompose the data using the K-d tree approach such that there
is no overlap in the partitioned data.
The partitioned data is then expanded to include ghost regions to the extent that
it still fits in memory.
Later, the overlapping areas between data blocks become regions to place the
splitting plane to repartition data such that each block gets an equal number of particles.
Their results demonstrated better load balance was achieved among processes without additional
costs of pre-processing and expensive communication.
Their results also demonstrate higher parallel efficiency.
However, their work made two crucial assumptions
1) an equal number of particles in data blocks might translate to equal work, and
2) the constrained K-d tree decomposition leads to an even distribution of particles.
These assumptions do not always hold practically.

In conclusion,
pre-processing works can achieve load balance with an additional cost for
\textit{parallelize over data}.
This cost goes up with large volumes of data.
The overall time for completing particle advection might not benefit from the
additional cost of pre-processing, especially when the workload is not compute-intensive.
Most solutions that rely on dynamic load balancing suffer from increased communication
costs or are affected by the distribution of particles and the complexity of the vector field.
The work proposed by Zhang et al.~\cite{zhang2017dynamic} is promising but still does not guarantee optimal load balancing.

%
\paragraph{\boit{Parallelize over particles}}
\label{sec:overparticles}
\textit{``Parallelize over particles''} is a paradigm for work distribution in
flow visualization where $M$ particles are distributed among $N$ processors.
Most commonly, the particle distribution is done such that each process
is responsible for computing the trajectories of $\frac{M}{N}$ particles.
Each process is responsible for the computation of streamlines for particles assigned to it.
This is done by loading the data blocks required by the process in order to advect the particles.
Particles are advected until they can no longer continue within the current data block,
in which case another data block is requested and loaded.

Previous works have explored different approaches to optimize the scheme described above.
Since the blocks of data are loaded whenever requested,
the cost of I/O is a dominant factor in the total time.
Prefetching of data involves predicting the next needed data block 
while continuing to advect particles in the current block to hide the I/O cost.
Most commonly, predictions are made by observing the I/O access patterns.
Rhodes et al.~\cite{rhodes2005iteration} used these access patterns as a priori
knowledge for caching and prefetching to improve I/O performance dynamically.
Akande et al.~\cite{akande2013iteration} extended their work to unstructured grids.
The performance of these methods depends on making correct predictions of the required blocks.
One way to improve the prediction accuracy is by using a graph-based approach
to model the dependencies between data blocks.
Some works used a preprocessing step to construct these graphs
~\cite{chen2011flow, chen2012flow, chen2013graph}.
Guo et al.~\cite{guo2014advection} used the access dependencies to produce
fine-grained partitions that could be loaded at runtime for better efficiency of data accesses.

Zhang et al.~\cite{zhang2016efficient} presented an idea of higher-order access transitions, 
which produce a more accurate prediction of data accesses.
They incorporated historical data access information to calculate access dependencies.

Since particles assigned to a single process might require access to
different blocks of data, most of the works using parallelization over particles 
use a cache to hold multiple data blocks.
The process advects all the particles that occur within the blocks of
data currently present in the cache.
When it is no longer possible to continue computation with the data in the cache,
blocks of data are purged, and new blocks are loaded into the cache.
Different purging schemes are employed by these methods,
among which ``Least-Recently Used,'' or LRU is most common.
Lu et al.~\cite{lu2014scalable} demonstrated the benefits of using a cache in their
work for generating stream surfaces.
They also performed a cache-performance trade-off study to determine
the optimal size of the cache.

Camp et al.~\cite{camp2010streamline} presented work comparing the MPI only and
MPI-hybrid implementations of parallelizing over particles.
Their objective was to prove the efficacy of using shared memory parallelism
with distributed memory to reduce communication and I/O costs.
They observed 2x-10x improvement in the overall time for calculation of streamlines
while using the MPI-hybrid version.

Along with caching, Camp et al.~\cite{camp2011evaluating} also presented work
that leveraged different memory hierarchies available on modern supercomputers
to improve the performance of particle advection.
The objective of the work is to reduce the cost of I/O operations.
Their work used Solid State Drives (SSDs) and local disks to store data blocks,
where SSDs are used as a cache.
Since the cache can only hold limited amounts of data compared to local disks,
blocks are purged using the LRU method.
When required blocks are not in the cache,
the required data is searched in local disks before accessing the file system.
The extended hierarchy allows for a larger than usual cache,
reducing the need to perform expensive I/O operations.

One trait that makes the parallel computation of integral curves challenging is
the dynamic data dependency.
The data required to compute the curve cannot be determined in advance unless there is a priori knowledge of the data.
However, this information is crucial for optimal load-balanced parallel scheduling.
One solution to this problem is to opt for dynamic scheduling.
Two well-studied techniques for dynamic scheduling are \textit{work-stealing}
and \textit{work-requesting}.
In both approaches, an idle process acquires work from a busy process.
Popularly, idle processes are referred to as thieves,
and busy processes are referred to as victims.
The major distinction between work-stealing and work requesting is how the thief acquires work from the victim.
In work-requesting, the thief requests work items, and the victim voluntarily shares it.
In work-stealing, the thief directly accesses the victim’s queue for work items
without the victim knowing.

A large body of works addresses \textit{work-stealing} in \textit{task-based}
parallel systems in general~\cite{blumofe1999scheduling,dinan2009scalable, shiina2019almost}.
In the case of integral curve calculation,
task-based parallelism inspires the parallelize over particles scheme. 
Dinan et al.~\cite{dinan2009scalable} demonstrated the scalability of the
work-stealing approach.
Lu et al.~\cite{lu2014scalable} presented a technique for calculating
stream surface efficiently using work-stealing.
%
%
Their algorithm aimed for the efficient generation of stream surfaces.
The seeding curve for streamlines was divided into segments,
and these segments were assigned to processes as tasks.
In their implementation, each process maintains a queue of segments.
When advancing the streamline segment using the front advancing algorithm
proposed by Garth et al.~\cite{garth2004surface}, if a segment
starts to diverge, it is split into two and placed back in the queue.
When a processor requires additional data to advance a segment,
it requests the data from the processes that own the data block.
Their solution demonstrated good load balancing and scalability.

Work stealing has been proven to be efficient in theory and practice.
However, Dinan et al. reported its implementation is complicated.

Muller et al.~\cite{muller2013distributed} presented an approach that used
work requesting for tracing particle trajectories.
Their algorithm started by equally distributing all work items (particles) among processes.
However, they started by assigning all particles to a single process for
performing the load balancing study.
Every time an active particle from the work queue
is unable to continue in the currently cached data,
it is placed at the end of the queue.
Whenever a thief tries to request work,
the particles from the end of the queue are provided,
reducing the current processes’ need to load the data block for the particle.
The results reported performance improvements between 30\% to 60\%.

Binyahib et al.~\cite{binyahib2019situ} compared the parallelize over particle
strategy to parallelize over data for its in-situ applicability.
Their findings suggest that for workloads where particles are densely seeded
in a certain region of the data, parallelize over partilces is a much better
strategy and can result in speedups upto $10\times$.

According to Childs et al.~\cite{childs2010extreme},
the dominant factor affecting the performance of parallelizing over particles is I/O.
The solution to solve the I/O problem during runtime is to perform prefetching of data.
However, works that propose prefetching incur additional costs of making predictions
of which blocks to read.
Leveraging the memory hierarchy similar to Camp et al. is a good strategy,
provided proper considerations for vector field size and complexity are made.
Apart from I/O costs, load balancing remains another factor affecting performance adversely.
Previous work stealing and work requesting strategies have demonstrated good
load balance with additional costs of communicating work items.
These costs could potentially be restrictive in the case of workloads with
a large number of particles.

\begin{table}
  \centering
  \begin{tabular}{lll}
    \toprule 
    \multirow{2}{*}{Problem Classification} & \multicolumn{2}{c}{Parallelization Strategy} \\[1ex]
    & Over Data & Over Particles \\
    \midrule
    Dataset size               & Large  & Small \\[1ex] 
    Number of particles        & Small  & Large \\[1ex]
    Seed Distribution          & Sparse & Dense \\[1ex]
    Vector Field Complexity    & No critical & No circular \\
                               & points      & field       \\ [1ex]
    \bottomrule
  \end{tabular}
  \caption{\label{tab:distsynth} Recommendation of parallelization strategy for
           particle advection workloads based on features of the problem.
           This table appears in the survey by Binyahib~\cite{binyahib2019scientific}.}
\end{table}
\begin{table*}
\centering
  \begin{tabular}{llrrrrlll}
    \toprule
    Algorithm & Architecture & Procs. & Data set & Time   & Seed  & Application& Seeding  &  Intent /  \\
              &              &        & Size     & steps  & Count &            & Strategy &  Evaluation      \\
    \midrule
Yu et al.                & Intel Xeon & 32  & 644M  & -  & 1M  & streamlines,& -  &  hierarchical    \\
\cite{yu2007parallel}    & (8x4)      & & & &                  & pathlines   &    &  representation, \\
                         & AMD Optron & 256 & 644M & 100 & 1M  &             & -  &  strong scaling  \\
                         & (2048x2)   & & & & & &  & \\[1ex]

Chen et al.               & Intel Xeon     & 32  & 162M      & -     & 700  & streamlines & -       & data partitioning /   \\
\cite{chen2008optimizing} & (48x2)         &     &           &       &      &             &         & strong scaling        \\[1ex]

Pugmire et al.              & Cray XT5 (ORNL)   & 512 & 512M & -   & 4k, 22K & streamlines  & uniform  & data loading,       \\
\cite{pugmire2009scalable}  & (149K)            & 512 & 512M & -   & 10K     &              & uniform  & data partitioning / \\
                            &                   & 512 & 512M & -   & 20K     &              & uniform  & weak scaling        \\[1ex]

Peterka et al.            & PowerPC-450    & 16k & 8B     & -   & 128k & streamlines, & random   & domain decomposition,      \\
\cite{peterka2011study}   & (40960x4)      & 32K & 1.2B   & 32  & 16k  & pathlines    & random   & dynamic repartitioning /   \\
                          &                &     &        &     &      &              &          & strong and weak scaling    \\[1ex]

Camp et al.                 & Intel Xeon  & - & 512M & - & 2.5K, 10K  & streamlines & dense,  & Effects of       \\
\cite{camp2011evaluating}   & Dash (SDSC) & - & 512M & - & 2.5K, 10K  &             & uniform & storage hierarchy\\
                            &             & - & 512M & - & 2.5K, 10K  &             &         &                  \\[1ex]
Camp et al.                 & Cray XT4 (NERSC) & 128 & 512M & - & 2.5K, 10K  & streamlines & dense,  & MPI-hybrid   \\
\cite{camp2010streamline}   & 9572x4           & 128 & 512M & - & 2.5K, 10K  &             & uniform & parallelism  \\
                            &                  & 128 & 512M & - & 1.5K, 6K   &             &         &              \\[1ex]
Nouanesengsy                      & PowerPC-450 & 4K  & 2B     & -  & 256K & streamlines & random  & workload aware             \\
et al.~\cite{nouanesengsy2011load}& (1024x4)    & 4K  & 1.2B   & -  & 128K &             & random  & domain decomposition /     \\
                                  &             &     &        &    &      &             &         & strong and weak scaling    \\[1ex]
Nouanesengsy                            & PowerPC-450  & 1k  & 8M     & 29  & 186M   & FTLE & uniform   & pipelined temporal         \\
et al.~\cite{nouanesengsy2012parallel}  & (40960x4)    & 1K  & 25M    & 48  & 65.2M  &      & uniform  & advection, caching /        \\
                                        &              & 16k & 345M   & 36  & 288M   &      & uniform  & strong and weak scaling     \\
                                        &              & 16K & 43.5M  & 50  & 62M    &      & uniform  &                             \\[1ex]
Camp et al.                & Cray XT4 (NERSC) & 128 & 512M & -  & 128 & stream surface & rake & Comparison of              \\
\cite{camp2012parallel}    & (9572x4)         & 128 & 512M & -  & 361 &                & rake & parallelization algorithms \\
                           &                  & 128 & 512M & -  & 128 &                & rake & for stream surfaces        \\[1ex]
Muller et al.                    & AMD Magny-Cours  & 1K & 32M  & 735 & 1M & streamlines, & uniform & work requesting / \\
\cite{muller2013distributed}     & (6384x24)        &    &      &     &    & pathlines    &         & load balancing,   \\
                                 &                  &    &      &     &    &              &         & strong scaling    \\[1ex]
Childs et al.              & Nvidia Kepler  & 8   & 1B & - & 8M & source-dest & uniform & Distriburted particle     \\
\cite{childs2014particle}  & (1 GPU / Proc) &     &    &   &    &             &         & advection over different  \\
                           & Intel Xeon     & 192 & 1B & - & 8M &             &         & hardware architectures /  \\
                           &                &     &    &   &    &             &         & comparison, strong scaling\\[1ex]
Guo et al.                 & Intel Xeon   & 64  & 755M          & 100 & -   & streak surface,& seed line & sparse data      \\
\cite{guo2014advection}    & (8x8)        & 64  & 3.75M         & 24  & 200 & pathlines,     & uniform   & management /     \\
                           & Intel Xeon   & 512 & 25M           & 48  & -   & FTLE           & uniform   & strong scaling   \\
                           & (700x12)     &     &               &     &     &                &           &                  \\[1ex]
Lu et al.             & PowerPC A2   & 1K & 25M  & - & 32K & stream surface & rakes & caching,        \\
\cite{lu2014scalable} & (2048x16)    & 4K & 80M  & - & 32K &                & rakes & performance /   \\
                      &              & 8K & 500M & - & 32K &                & rakes & strong scaling  \\
                      &              & 8K & 2B   & - & 64K &                & rakes &                 \\[1ex]
Zhang et al.               & Intel Xeon  & 64  & 3.75M   & 24  & 6250 & pathlines & uniform & data prefetching / \\
\cite{zhang2016efficient}  & (8x8)       & 64  & 25M     & 48  & 4096 &           & -       & strong scaling     \\[1ex]
Zhang et al.            & PowerPC A2 & 8K  & 1B     & -   & 128M & streamlines, &  -      & domain decomposition,   \\
\cite{zhang2017dynamic} & (2048x16)  & 8K  & 3.8M   & 24  & 8M   & source-dest, &  -      & using K-d trees /       \\
                        &            & 8K  & 25M    & 48  & 24M  & FTLE         & uniform & strong and weak scaling \\[1ex]
Binyahib et al.            & Intel Xeon & 512 & 67M & - & 1M  & source-dest & dense,  & In situ parallelization \\
\cite{binyahib2019situ}    & (2388x32)  &     &     &   &     &             & uniform & over particles          \\[1ex]
Binyahib et al.            & Intel Xeon & 1K         & 34B & - & 343M  & source-dest & dense,  & Comparison of              \\
\cite{binyahib2020bakeoff} & (2388x32)  & (8K cores) &     & - &       &             & uniform & parallelization algorithms \\[1ex]
Binyahib et al.            & Intel Xeon & 1K         & 34B & - & 343M  & source-dest & dense,  & novel hybrid  \\
\cite{binyahib2021hylipod} & (2388x32)  & (8K cores) &     & - &       &             & uniform & parallelization algorithm\\[1ex]
\bottomrule
\end{tabular}
\caption{Summary of large scale distributed particle advection worklets.
The numbers in parenthesis in the Architecture column represent the total number of cores
available on the execution platform.
\label{tab:distscale}
}
\end{table*}
\begin{table}
\centering
\begin{tabular}{lr}
\toprule
Application & Particles /1k Cells \\
\midrule
Souce-destination & 72222.20 \\[1ex]
FTLE              & 5013.02  \\[1ex]
Streamlines       & 9.89     \\[1ex]
Pathlines         & 6.93     \\[1ex]
Stream surface    & 0.25     \\[1ex]
\bottomrule
\end{tabular}
\caption{Number of particles used per one thousand cells of data for different
applications from works described in Table~\ref{tab:distscale}.
\label{tab:partpercell}
}
\end{table}

\begin{figure*}
\centering
\includegraphics[width=0.75\linewidth]{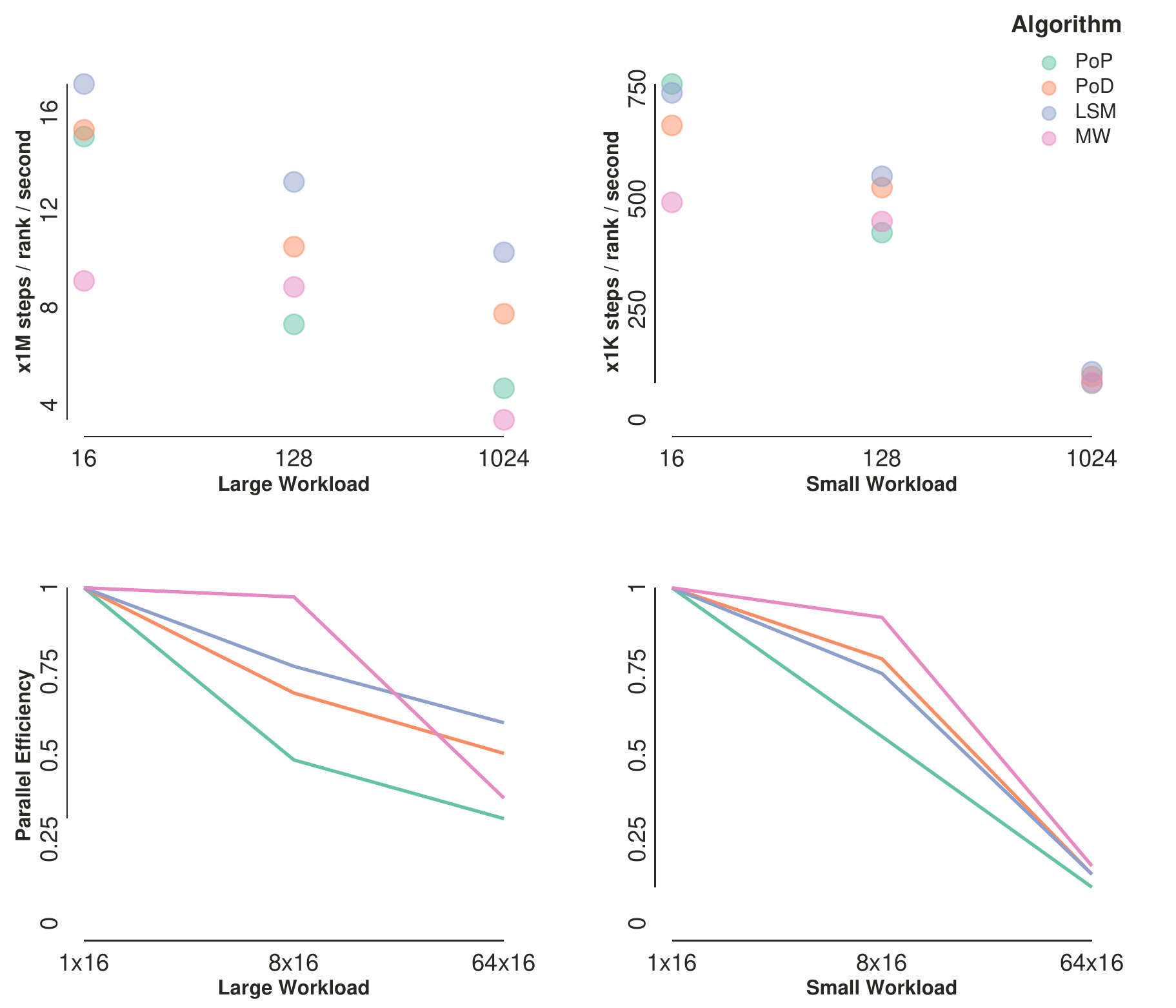}
\caption{Weak scaling plots for distributed memory particle advection based on
the comparison of four parallelization algorithms by Binyahib et al.~\cite{binyahib2020bakeoff}.
The plots present performance comparison of the algorithms for two different workloads.
The large workload used 1 particle per 100 cells where each particle advanced 10K steps.
The small workload used 1 particle per 10K cells where each particle advanced 1K steps.
The plots on the top present the performance of the algorithms in terms of throughput along the Y axis,
while the plots on the bottom present the parallel efficiency while using weak scaling along the Y axis.
In all cases the X axis represents the number of MPI ranks used to perform the experiments.
\label{fig:distsum}}
\end{figure*}
\paragraph{\boit{Hybrid Particle Advection}}
\label{sec:hybridadv}
The works described in this section combine \textit{parallelize over data}
and \textit{parallelize over particles} schemes to achieve optimal load balance.
Pugmire et al.~\cite{pugmire2009scalable} introduced an algorithm that uses a
master-worker model.
The processes were divided into groups, and each group had a master process.
The master is responsible for maintaining the load balance between processes as
it coordinates the assignment of work.
The algorithm begins with statically partitioning the data.
All processes load data on demand.
Whenever a process needs to load data for advancing its particles,
it coordinates with the master.
The master decides whether it is more efficient for the process to load data
or to send its particles to another process.
The method proved to be more efficient in I/O and communication than the traditional
parallelization approaches.

Kendall et al.~\cite{kendall2011simplified} provided a hybrid solution
which they call DStep and works like the MapReduce framework~\cite{dean2008mapreduce}.
Their algorithm used groups for processes as well and has a master
to coordinate work among different groups.
A static round-robin partitioning strategy is used to assign data blocks to processes,
similar to Peterka et al.~\cite{peterka2011study}.
The work of tracking particles is split among groups where the master process
maintains a work queue and assigns work to processes in its group.
Processors within a group can communicate particles among them.
However, particles across groups can only be communicated by the master processes.
The algorithm provided an efficient and scalable solution for particle tracing
and has been used by other works~\cite{guo2013coupled, guo2014scalable,liu2016comparative}.

Binyahib et al~\cite{binyahib2021hylipod} proposed a new `HyLiPoD' algorithm for particle advection.
Their work was inspired from the finding of the previous bake-off study comparing
different distributed particle advection strategies~\cite{binyahib2020bakeoff}.
HyLiPoD is short for Hybrid Lifeline and Parallelize over Data,
and the algorithm aims to choose the best strategy between the Lifeline algorithm~\cite{binyahib2019lifeline}
and parallelize over data for distributed particle advection given a certain workload. 
%
%
%
%
%
%
%

%
\subsubsection{Summary}
\label{sec:distsummary}
This section summarizes distributed particle advection in two parts.
First, general take-aways are discussed based on the various factors discussed
in the introduction of this section.
Second, observations from the studies in terms of their
particle advection workloads are presented.

Table \ref{tab:distsynth} provides a simple lookup for a parallelization strategy
based on various workload factors discussed earlier in the section.
These strategies were presented in a survey by Binyahib~\cite{binyahib2019scientific}.
\boit{Parallelize over data} is best suited when the data set volume is large.
However, in the presence of flow features like critical points and vortices,
parallelize over data can suffer from load imbalance.
While several methods have been proposed for data repartitioning for load-balanced computation,
these works incur the cost of pre-processing and redistributing data.
\boit{Parallelize over particles} is best suited when the number of particles is large.
It can suffer from load imbalance due to inconsistencies in the computational work
for different particles.
Some works aim to address the problem of load imbalance but have added costs of pre-processing,
communication, and I/O.
\boit{Hybrid} solutions demonstrate better scalability and
efficiency compared to the traditional methods.
However, implementing these methods is very complicated and typically has some
added cost of communication and I/O.

Figure \ref{fig:distsum} shows a comparison of scaling behaviors of four 
parallelization algorithms, extracted from the study presented by Binyahib et al.~\cite{binyahib2020bakeoff}.
These algorithms include parallelize over particles, parallelize over data,
Lifeline Scheduling Method (LSM, an extension of parallelize over particles)~\cite{binyahib2019lifeline},
and master-worker (a hybrid parallel algorithm).
The Figure presents a weak scaling of these algorithms.
The top row plots show the throughput of these algorithms in terms of
number of steps completed by each MPI rank per second.
The bottom row plots show the efficiency of weak scaling achieved by the
different algorithms.
The efficiency of the algorithms drop significantly as the concurrency and
workload are increased. 
The drop is more significant in smaller workloads than in larger workloads.
The only study which compared the scaling behaviors of the most widely used
parallelization algorithm used weak scaling.
In order to be able to quantify the speed-ups resulting from added distributed
parallelism for a given workload, a strong scaling study is necessary.
The strong scaling study for these algorithms is a potential avenue for future research.

Table \ref{tab:distscale} summarizes large-scale parallel particle advection-based
flow visualization studies in terms of the distributed executions and the magnitudes of the workloads. 
The platforms used by the considered studies in this section span from
desktop computers to large supercomputers.
The work with the least amount of processes and workload in this survey is by Chen et al.~\cite{chen2008optimizing},
which used only 32 processes to produce 700 streamlines. 
The work with the largest number of processes was by Nouanesengsy et al.~\cite{nouanesengsy2012parallel},
which used 16 thousand processes for FTLE calculation.
However, the work with the most workload was by Binyahib et al.~\cite{binyahib2020bakeoff}, 
which used 343 million particles for advection in data with 34 billion cells.

Table \ref{tab:partpercell} summarizes the number of particles used in proportion
to the size of the data used in the works included in Table \ref{tab:distscale}. 
Stream surface generation is the application that required the least amount of particles.
A significant part of the cost of generating stream surfaces comes from triangulating
the surfaces from the advected streamlines. 
These streamlines cannot be numerous as they may lead to issues like occlusion.
Source-destination queries use the most particles in proportion to the data size.
All other applications need to store a lot of information in addition to the 
final location of the particle ---
streamlines and pathlines need to save intermediate locations for representing the trajectories,
stream surfaces need the triangulated surface for rendering,
and FTLE analysis needs to generate an additional scalar field.
Source-destination analysis has no such costs and can instead use the savings
in storage and computation to incorporate more particles. 
%

\subsection{Hybrid Parallelism for Particle Advection}
\label{sec:hybrid}

Hybrid parallelism refers to a combination of using shared- and distributed-memory parallel techniques.
For these works, the distributed-memory elements managed dividing work among nodes,
and the shared-memory parallelism approach was providing a ``pool'' of cores that could advect particles quickly.
Camp et al.\cite{camp2010streamline}
presented two approaches that used
multi-core processors to parallelize particle advection
1) parallelization over particles, and 2) parallelize over data blocks.
In both cases, the authors aimed to use the N allocated cores.
For parallelization over particles,
N worker threads were used along with $N$ I/O threads.
The worker threads are responsible for performing particle advection.
The I/O threads manage the cache of data blocks to support the worker threads.
For parallelization over data blocks,
$N - 1$ worker threads are used, which access the cache of data blocks directly,
and an additional thread was used for communicating results with other processes.

Camp et al.~\cite{camp2013gpupa} also extended their previous work to GPUs.
One of their objectives was to compare particle advection
performance on the GPU against CPU under different workloads.
They varied the datasets,
the number of particles,
and the duration of advection for their experiments.
Their findings suggest that in the case where the workloads have fewer particles
or longer durations, the CPU performed better.
However, in most other cases, the GPU was able to outperform the CPU.

Childs et al.~\cite{childs2014particle} explored particle advection performance
across various GPUs (counts and device) and CPUs (processors and concurrency).
Their objective was to explore the relationship between parallel device choice
and the execution time for particle advection.
Two of their key findings were:
1) For CPUs, adding more cores benefited workloads that execute for medium to longer duration,
2) CPUs with many cores were as performant as GPUs and often outperformed GPUs for small workloads with short execution times.
3) With higher particle densities ($50^3$ or more) GPUs can be saturated and result in performance imporvements
proportpional to their FLOP rates, faster GPUs can provide better speedups.

Jiang et al.~\cite{jiang2013} studied shared memory multi-threaded generation of streamlines with a locally attached NVRAM.
Their particular area of interest was in understanding data movement strategies that will keep the threads busy performing particle advection.
They used two data management strategies.
The first used explicit I/O to access data.
The second was a kernel-managed implicit I/O method that used memory-mapping to provide access to data.
Their study indicated that thread oversubscription of streamline tasks is an effective method for hiding I/O latency,
which is bottleneck for particle advection.

\section{Conclusion and Future Work}
\label{sec:conclusion}

\begin{figure}
\centering
\includegraphics[width=1.0\linewidth]{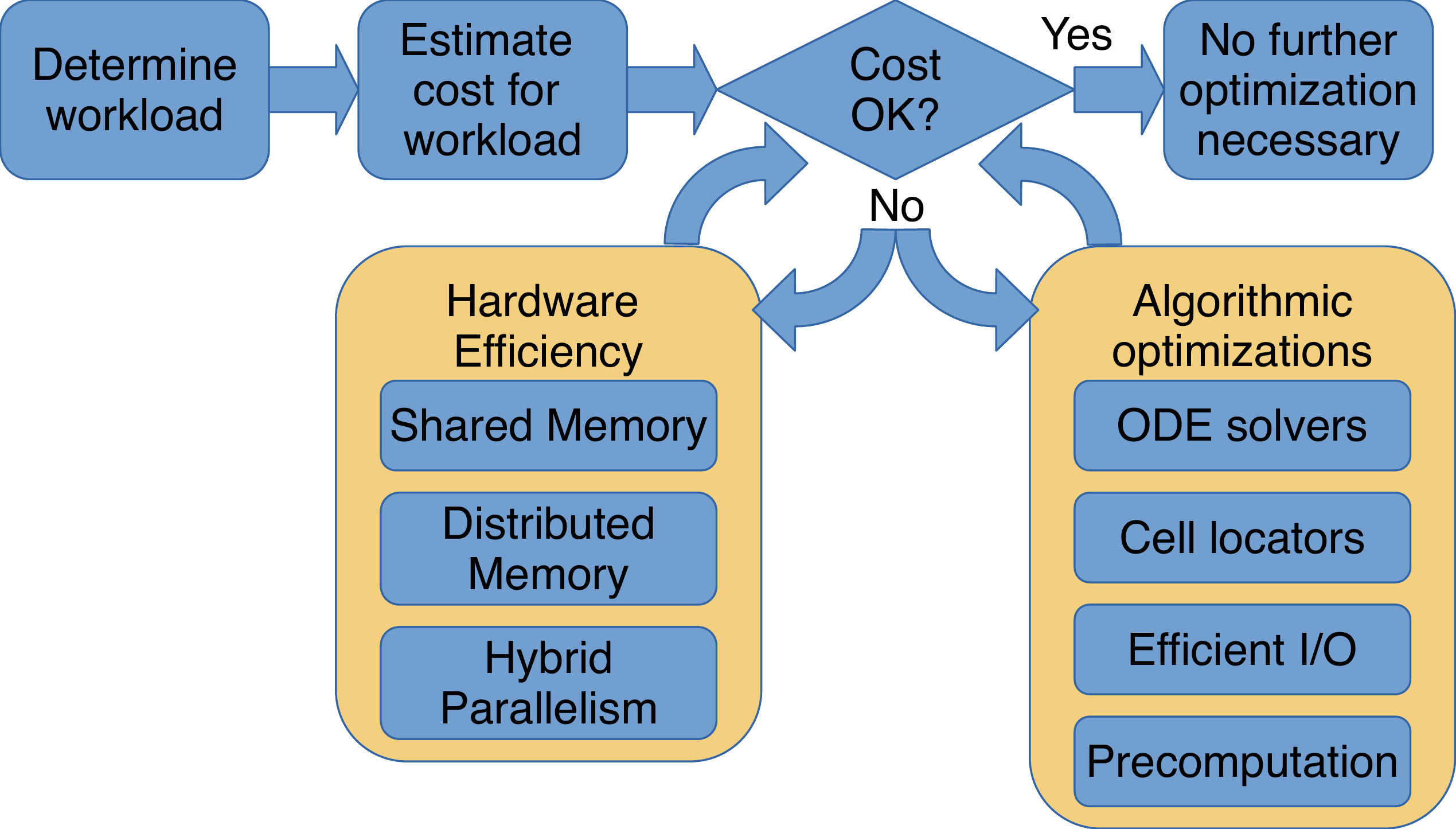}
\caption{\label{fig:decisions} A flow chart to determine the potential optimizations
to be applied to a flow visualization algorithm.}
\end{figure}

This survey has provided a guide to particle advection performance.
The first two parts focused on high-level concepts:
particle advection building blocks and a cost model.
The last two parts surveyed existing approaches for algorithmic optimizations 
and parallelism.
While the guide has summarized research findings to date, additional
research can make the guide more complete.
Looking ahead to future work,
we feel this survey has illuminated three types of gaps in the
area of particle advection performance.

The first type of gap involves the lack of holistic studies to inform behavior across
diverse workloads.
Adaptive step sizing, since its focus is more on accuracy than performance,
can lead to highly varying speedups.
Understanding when speedups occur and their magnitude would be very helpful
for practitioners when deciding whether to include this approach.
Similarly, the expected speedup for a GPU is highly varied based on workload
and GPU architecture.
While this survey was able to synthesize results from a recent study~\cite{pugmire2018portable}, significantly more detail would be useful.

The second type of gap covers possible optimizations that have not yet been pursued.
All of the hardware efficiency works in this survey involved parallelism,
yet there are still additional hardware optimizations available.
In the ray tracing community --- similar to particle advection in that rays
move through a volume in a data-dependent manner ---
packet tracing,
where rays on similar trajectories are traced together,
has led to significant speedups.
Further, there can be significant improvement from complex schemes.
For example, Benthin et al.~\cite{6081859} employed a hybrid approach that generates and traces rays in packets and then automatically switches to tracing rays individually when they diverge.  This hybrid algorithm outperforms conventional packet tracers by up to 2X.
Finally, there are additional types of optimizations.
Taking another example from ray tracing,
Morrical et al.~\cite{morrical_rtxpoints_2020} presented a method that improved the performance of direct unstructured mesh point location~\cite{Sawhney2020} by using Nvidia RTX GPU.
Their approach re-implemented the point location problem as a ray tracing problem, which enabled tracing the points using the hardware.
Their results showed equal or better performance compared to state-of-the-art solutions and could provide inspiration for improved cell locators on GPUs for particle advection.

The third type of gap is in cost modeling.
While our cost model is useful in the context of providing a guide
for particle advection performance, 
future work could make the model more powerful.
One possible extension relates to the first type of gap (lack of holistic studies).
In particular, holistic studies on expected speedups over diverse workloads
would enable adaptive step sizing and GPU acceleration to fit within the model.
Another  possible optimization is to broaden the model.
In particular, I/O is not part of our cost model, so optimizations for
I/O efficiency cannot be included at this time.
Further, precomputation likely requires a different type of model altogether,
i.e., using a form of the current model for regular particle advection and
a different model for precomputation, and selecting the best approach from the two.
Finally, distributed-memory parallelism is often applied to very large data sets
that cannot fit into memory of a single node,
which significantly increases modeling complexity.
%
That said, these extensions could have significant benefit.
One benefit would be using prediction to adapt workloads to fit available
runtime.
A second (perhaps more powerful) benefit would be to enable a workflow for
decision-making.
This workflow is shown via a flow chart in Figure \ref{fig:decisions},
and would operate in three steps.
In the first step, the desired workload would be analyzed to see how many operations need to be performed.
In the second step, the analysis from the first step would be used to estimate
the execution time costs to execute the algorithm.
In the third step, the estimated costs from the second step would be
compared to user requirements.
If the estimated costs are within the user's budget,
then no optimizations are necessary and the workload can be executed as is.
If not, then candidate optimizations should be considered 
and the workflow should be repeated with candidate optimizations 
until the desired runtime is predicted.


%

\ifCLASSOPTIONcompsoc
  \section*{Acknowledgments}
\else
  \section*{Acknowledgment}
\fi

\ifCLASSOPTIONcaptionsoff
  \newpage
\fi



\bibliographystyle{IEEEtran}
\bibliography{advection-tvcg,surveys,algorithmic,precomputation,shared,gpuadvection,distributed,hw-optimization}

\appendices

\section{Evaluating Particle Advection Performance Cost Model}
\label{app:costmodel}
This appendix further evaluates our cost model for particle advection performance 
in three parts.
First, it considers the actual costs for terms in the 
cost model, measured in floating point operations.
Second, it
considers a notional example of how the cost model can be used
to predict the number of floating point operations.
Finally, it provides an evaluation of the cost model's
accuracy on various workloads, and also provides an approach for converting
the output of the cost model to execution time.

\subsection{Evaluating Cost Model Terms in Floating-Point Operations}
\label{sec:real_values}

\begin{table*}
\centering
\begin{tabular}{lrrrrrr}
\toprule
Solver  & Data set     & $solve$ & $locate$ & $interp$ & $terminate$ & Total \\
        & type         &         &          &          &             &       \\
\midrule
Euler   & Uniform      &       6 & 15       & 15       & 5            &  41  \\
        & Rectilinear  &       6 & 17       & 15       & 5            &  43  \\
        & Unstructured &       6 & 918      & 35       & 5            & 964  \\
RK4     & Uniform      &      37 & 15       & 15       & 5            & 162  \\
        & Rectilinear  &      37 & 17       & 15       & 5            & 170  \\
        & Unstructured &      37 & 918      & 35       & 5            & 3854 \\
\bottomrule
\end{tabular}
\caption{
\label{tab:costpred}
Analytical cost calculation for particle advection.
The costs in the table are by reference of a $50^3$ grid.
Hence, for the next two equations, $d=50$.
Rectilinear location costs : $3 \times log(d)$.
Unstructured location costs : $log(d^3) \ times 10 + 748$.
The costs for unstructured grid are highlighted in red as these are estimates based on a tree structure.
This study assumes $10$ FLOPs for checking each level of the tree.
An additional $748$ FLOPs are required to check if the point indeed
belongs inside the identified containing cell which is estimated using the Newton's method.
Each iteration of the Newton's method requires 374 FLOPs (code reviewed from VisIt),
and based on our experimental validation each check required 2 iterations to converge.
}
\end{table*}

\begin{figure*}[h]
\centering
\includegraphics[width=0.65\linewidth]{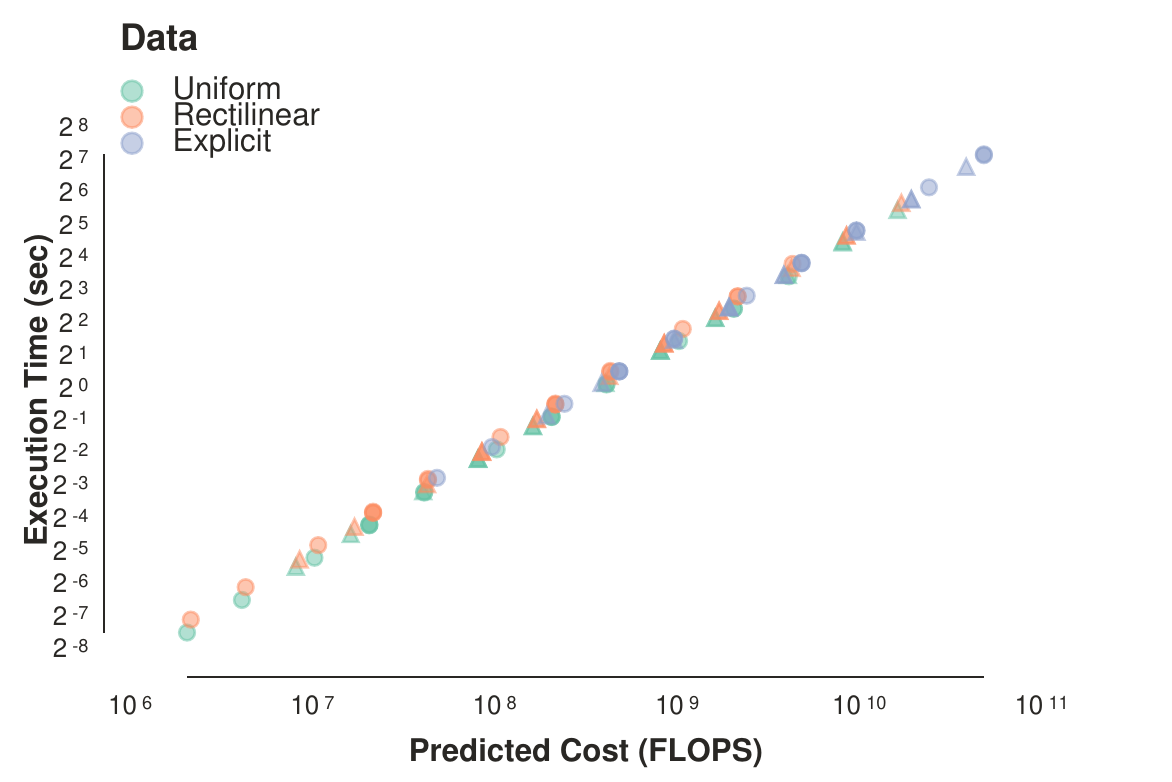}
\caption{
Comparing analytical cost and actual execution cost for particle advection.
The first four columns describe the workload.
The X axis represents the estimated number of FLOPS  for a workload using information
from Table \ref{tab:costpred} and our cost formula (Equation \ref{eq:complete}).
The Y axis represents the execution time of
running this workload on a single core using VTK-m.
All experimental points being collinear would build confidence
that an analytic approach can be used to estimate runtimes.
\label{fig:costfactor}
}
\end{figure*}

Equation~\ref{eq:complete} does not specify the unit of measurement for each
term and for the overall cost.
While time would be a common choice, we choose our unit of measurement to
be number of floating-point operations.
We make this choice since floating-point operations are consistent over architecture and 
caching effects, and, further, are easy to quantify (either by studying code or through profilers).
Finally, looking ahead to validation,
counting floating-point operations does track
closely with actual execution.

Table \ref{tab:costpred} presents the floating-point costs for the terms in Equation~\ref{eq:complete}
for a variety of instantiations:
RK4 and Euler solvers and three commonly used types of meshes.
%
The terms is this table were determined by studying code and counting floating-point operations.
As noted in the table caption, some operations vary based on mesh size and the table entries correspond
to a typical size.
Finally, cell location with unstructured meshes incorporate Newton's method, and we performed experiments
to find the average number of iterations to converge (2 iterations).

\subsection{Notional Usage of the Cost Model}
\label{sec:notional_usage}
This section demonstrates cost estimation for a hypothetical
workload based on Equation~\ref{eq:complete} and Table~\ref{tab:costpred}.
The hypothetical workloads consists of a flow visualization algorithm advancing 
a million particles in a 3D uniform grid for a maximum of 1000 steps
using a RK4 solver.
Then, the cost of each component can be estimated as follows:
\begin{itemize}
\item[] \boit{Locate:}
The locate operation in a uniform grid uses $15$ FLOP to find the cell
and the particle's location within the cell for interpolation.  
\item[] \boit{Interpolate:}
The interpolate operation in a uniform grid requires trilinear interpolation
which uses $15$ FLOP to evaluate the velocity at the given location.
\item[] \boit{Solve:}
The solve operation for the RK4 integration scheme uses $37$ FLOP to calculate
the determine the next position of the particle.
\item[] \boit{Analyze:}
The cost of analysis of the step varies based on the visualization technique being used.
In case only deals with advancing particles and hence the analysis cost is $0$.
\item[] \boit{Terminate:}
The terminate operation requires $5$ FLOP to determine if the particle is outside
the spatio-temporal bounds or if the particles completes the maximum number of steps.
\end{itemize}
These costs can be substituted in Equation~\ref{eq:complete} to get the final cost
of a single step of a particle in the presented situation.

\begin{equation}
\label{eq:costex}
\begin{split}
Cost & = \sum_{i=0}^{i=1M} \sum_{j=0}^{j=1000} \Big( 37 + \sum_{k=0}^{k=4} \big( 15 + 15 \big) + 0 + 5 \Big) \\
& = \sum_{i=0}^{i=1M} \sum_{j=0}^{j=1000} \Big( 37 + 120 + 0 + 5 \Big) \\
& = \sum_{i=0}^{i=1M} \sum_{j=0}^{j=1000} 162 \\
& = \sum_{i=0}^{i=1M} 162,000 \\
& = 162,000,000,000 \text{ FLOP}\\
\end{split}
\end{equation}

\subsection{Empirical Validation}
\label{sec:evaluation}

This section performs experimental validation of our cost model (Equation~\ref{eq:complete})
incorporating our measurements for the number of floating-point operations per term (Table~\ref{tab:costpred}).
It presents a set of experiments performed where the calculated cost
is translated into an execution time for a workload and them compared against
its actual execution time.
These experiments were performed on an Intel Xeon E5-1650 CPU with a clock rate
of 3.80 GHz using a particle advection implementation
from the VTK-m visualization library~\cite{moreland2016vtk} with a single CPU core.
%
The data used for the experiments was of the resolution $50\times50\times50$,
which matches the assumptions in Table~\ref{tab:costpred}.

The results of the experiments are presented in Figure~\ref{fig:costfactor}.
This figure shows a very strong fit between the predicted cost in FLOPS and the actual
execution time; statistical analysis shows a correlation coefficient of \fix{X}, and
a best fit line of \fix{Y=mx+b}.
In effect, the best fit line provides the actual time prediction.
For example, a workload with $10^8$ FLOPS would take \fix{$m\times 10^8+b$} seconds.

In the context of our workflow, a visualization application developer may choose to do more work.
First, they could run several experiments on their intended architecture to calculate their own
best fit line.
Second, they could recalculate Table~\ref{tab:costpred} using their own implementation and/or anticipated
data sizes.
That said, performing such additional work is likely unnecessary.
Our model and workflow are intended to infer coarse trends, and repeating our
analysis with new implementations, architectures, or data sets would likely not
yield a significantly different cost model.

\begin{IEEEbiography}
[{\includegraphics[width=1in,height=1.25in,clip,keepaspectratio]{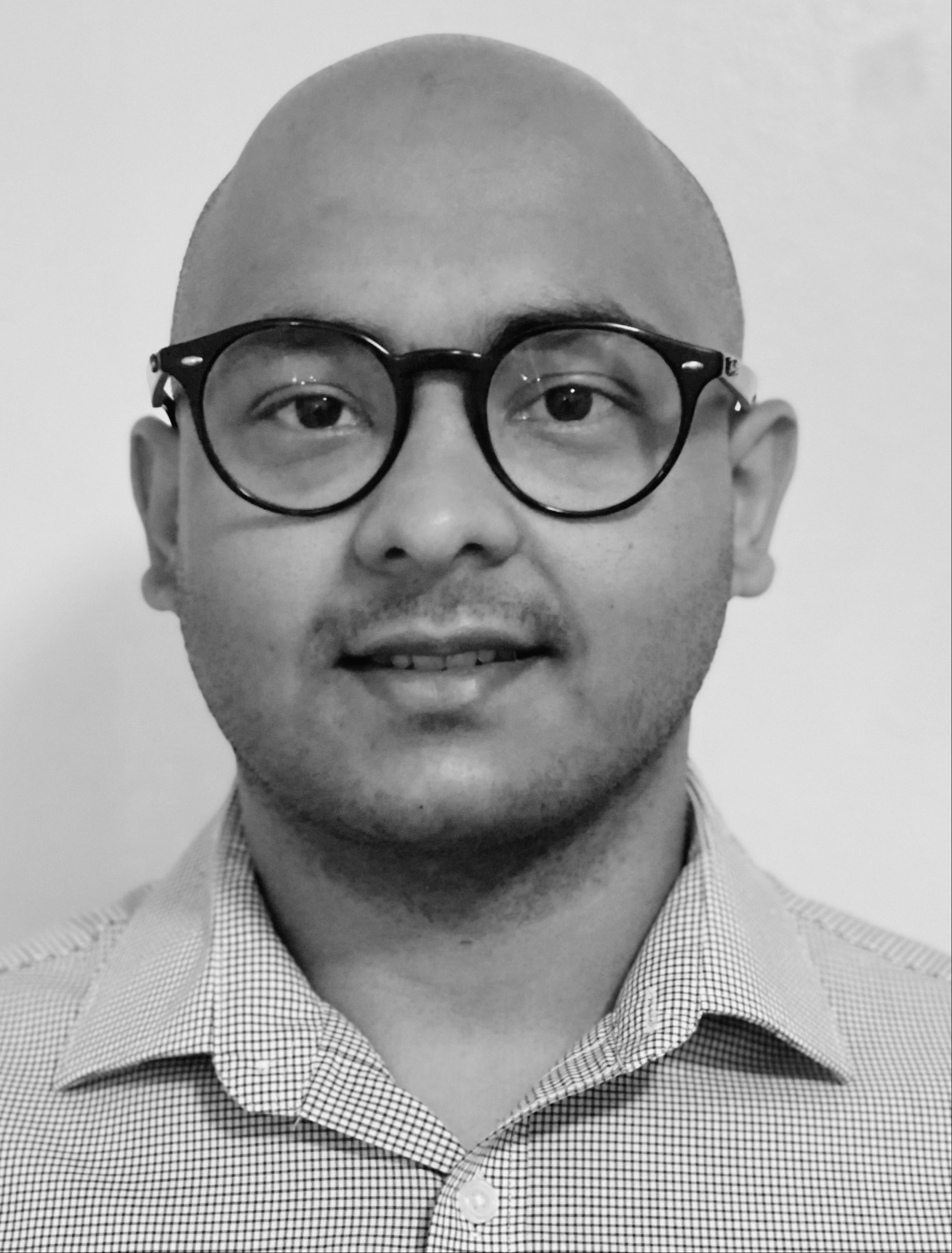}}]
{Abhishek Yenpure} is a Ph.D. candidate at the University of Oregon and CDUX research group led by Hank Childs. He received his Bachelor of Engineering degree from the University of Pune. His research interests include high performance computing and scientific visualization.
\end{IEEEbiography}
\begin{IEEEbiography}
[{\includegraphics[width=1in,height=1.25in,clip,keepaspectratio]{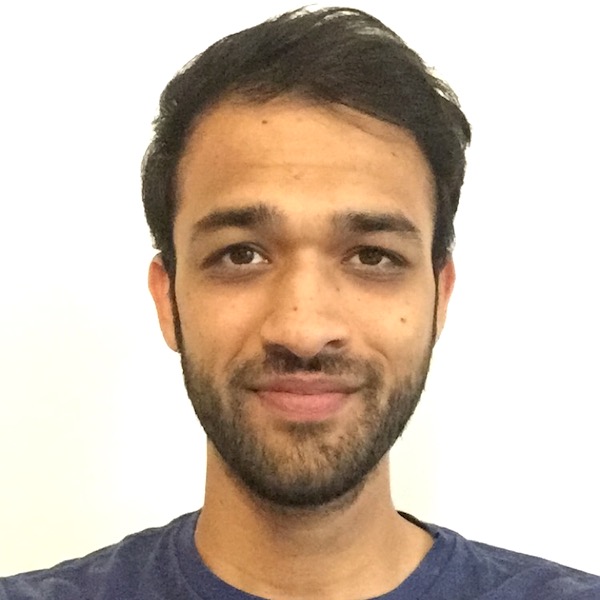}}]
{Sudhanshu Sane} completed a Ph.D. in Computer Science at the University of Oregon in June 2020. Upon graduation, Sudhanshu joined the Scientific Computing and Imaging Institute at the University of Utah as a postdoctoral research fellow. As of 2022, Sudhanshu is a software engineer at Luminary Cloud. Sudhanshu's research interests include computational fluid dynamics, uncertain and multivariate data visualization, and high performance computing. 
\end{IEEEbiography}
\begin{IEEEbiography}
[{\includegraphics[width=1in,height=1.25in,clip,keepaspectratio]{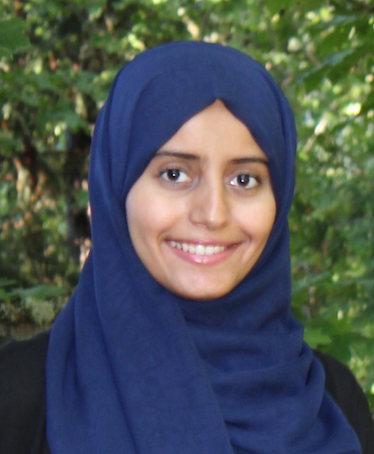}}]
{Roba Binyahib} is a graphics software engineer in the Advanced Rendering and Visualization Team at Intel.
She received her Ph.D. in computer science from the University of Oregon in 2020.
Her research is in the area of scientific visualization and high-performance computing,
focusing on developing visualization for large-scale HPC platforms.
\end{IEEEbiography}
\begin{IEEEbiography}
[{\includegraphics[width=1in,height=1.25in,clip,keepaspectratio]{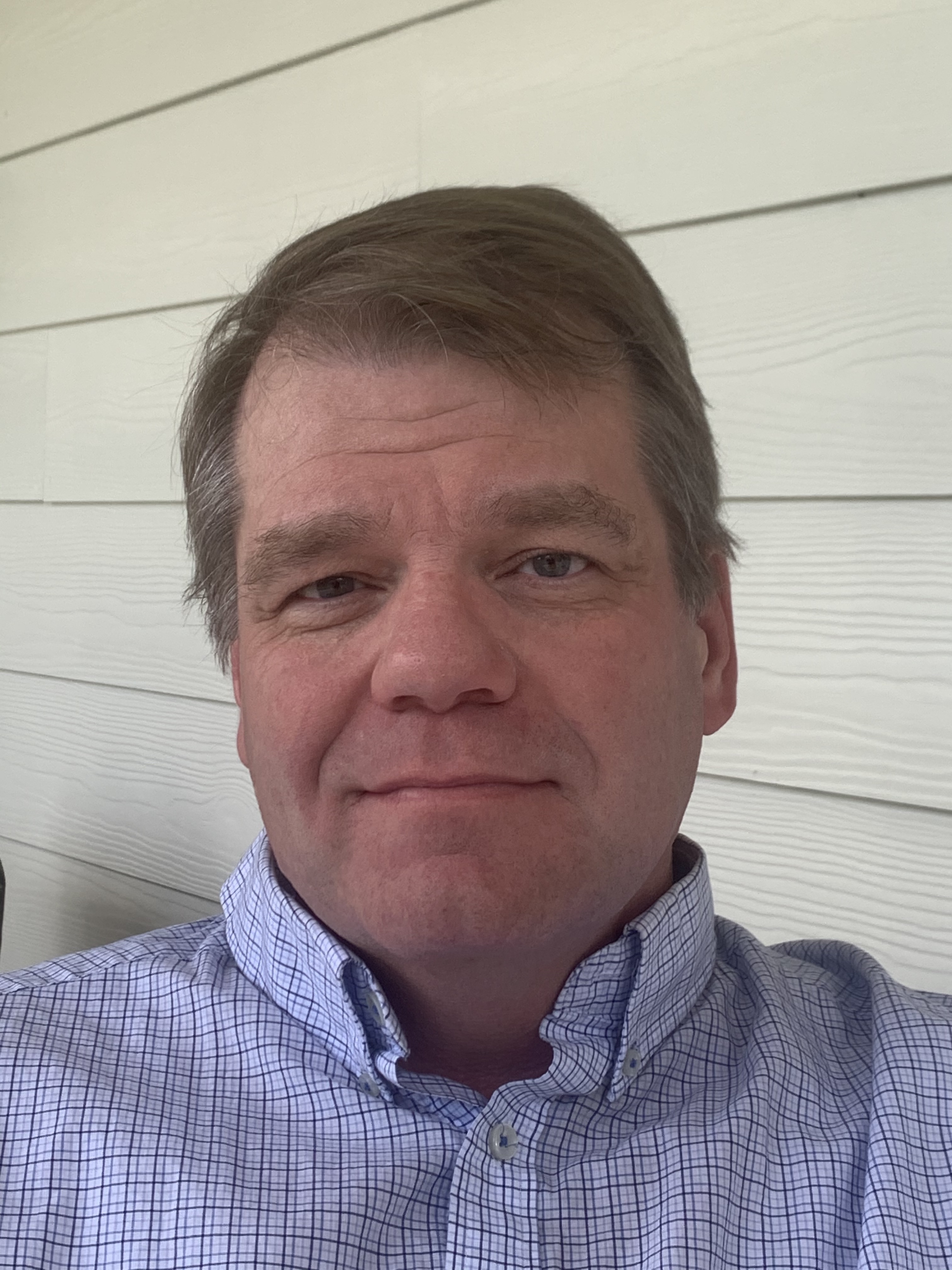}}]
{David Pugmire} is a distinguished scientist at Oak Ridge National Laboratory and a Joint Faculty Professor in the Electrical Engineering and Computer Science Department at the University of Tennessee. He received his Ph.D. in computer science from the University of Utah in 2000. His research interests are in large-scale parallelism for analysis and visualization of scientific data.
\end{IEEEbiography}
\begin{IEEEbiography}
[{\includegraphics[width=1in,height=1.25in,clip,keepaspectratio]{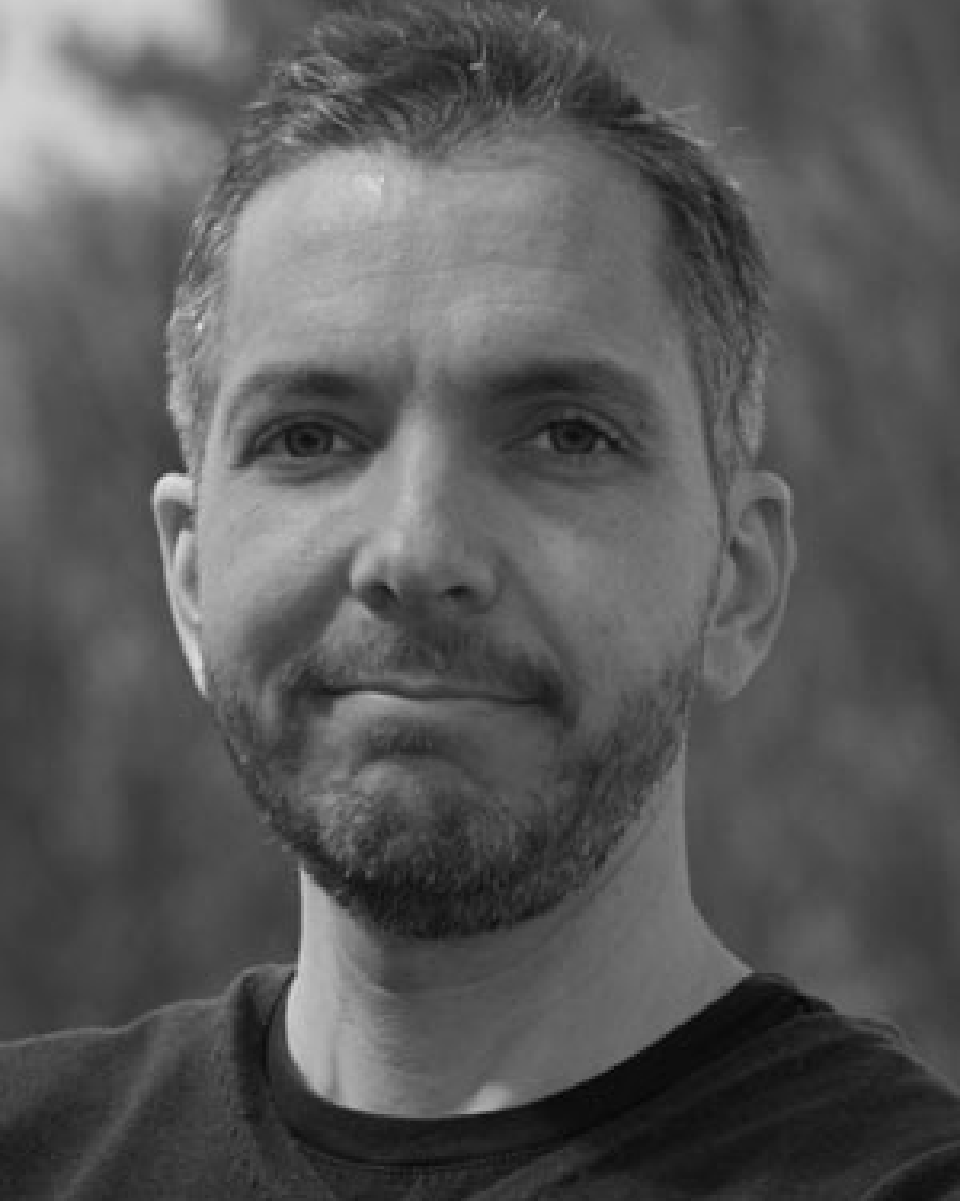}}]
{Christoph Garth} received the PhD degree in computer science from Technische Universit\"{a}t (TU) Kaiserslautern in 2007. After four years as a postdoctoral researcher with the University of California, Davis, he rejoined TU Kaiserslautern where he is currently a full professor of computer science. His research interests include large-scale data analysis and visualization, in situ visualization, topology-based methods in visualization, and interdisciplinary applications of visualization.
\end{IEEEbiography}
\begin{IEEEbiography}
[{\includegraphics[width=1in,height=1.25in,clip,keepaspectratio]{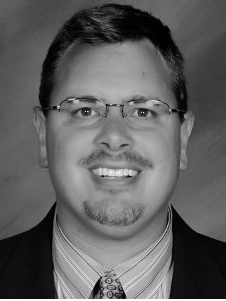}}]
{Hank Childs} is a professor of computer and information science at the University of Oregon.  He received a Ph.D. in computer science from the University of California at Davis in 2006. His research focuses on scientific visualization, high performance computing, and the intersection of the two.
\end{IEEEbiography}

\end{document}